\documentclass[aps, prb, superscriptaddress, twocolumn, letterpaper, 10pt, amsfonts, amsmath, amssymb]{revtex4-2}

\usepackage{graphicx}
\usepackage{hyperref} 
\usepackage{xcolor}
\usepackage{soul}
\hypersetup{
    colorlinks,
    linkcolor={blue!80!black},
    citecolor={blue!80!black},
    urlcolor={blue!80!black}
}
\usepackage{physics}
\usepackage{comment}
\usepackage{siunitx}
\usepackage{placeins}
\newcommand{\pdagger}{{\phantom{\dagger}}}

\usepackage{siunitx}
\usepackage{chemformula}

\usepackage{xcolor}
\usepackage{lipsum}

\usepackage{titletoc}

\begin{document}
\title{Intrinsic superconducting diode effect and nonreciprocal superconductivity in rhombohedral graphene multilayers}

\author{Yinqi Chen}
\affiliation{Hearne Institute of Theoretical Physics, Department of Physics \& Astronomy, Louisiana State University, Baton Rouge LA 70803, USA}

\author{Mathias S. Scheurer}
\affiliation{Institute for Theoretical Physics III, University of Stuttgart, 70550 Stuttgart, Germany}

\author{Constantin Schrade}
\affiliation{Hearne Institute of Theoretical Physics, Department of Physics \& Astronomy, Louisiana State University, Baton Rouge LA 70803, USA}

\date{\today}

\begin{abstract}
Recent experiments have revealed that superconductivity in rhombohedral tetralayer graphene can emerge from a valley-polarized and, hence, chiral normal state. The interplay of pairing and the reduced normal-state symmetries sparked widespread interest. In this work, we demonstrate within a microscopic theoretical formalism that this stabilizes a non-reciprocal superconducting state. Our results are based on a fully self-consistent framework for determining the superconducting order parameter from a Kohn-Luttinger mechanism. We show that the system displays a sizeable intrinsic superconducting diode effect, i.e., without the need for applying external magnetic fields, which is also highly tunable by an external displacement field. Moreover, we find that the angular dependence of the critical current is highly sensitive to the Fermi surfaces of the normal state. Hence, our results suggest that the critical current could provide insights into the type of Fermi surface topology from which superconductivity arises.
\end{abstract}

\maketitle

Non-reciprocal phenomena in superconducting systems that break time-reversal and inversion symmetry have recently attracted a great deal of attention~\cite{nadeem2023superconducting}.
A key example is the superconducting diode effect (SDE), characterized by a critical current that depends on the current-bias direction~\cite{ando_observation_2020,he_phenomenological_2022,yuan_supercurrent_2022,zhang_general_2022,misaki_theory_2021,davydova_universal_2022,souto2024tuning}.
One route to achieving such a SDE relies on applying an external magnetic field~\cite{zazunov2009anomalous,brunetti2013anomalous,pal2022josephson,baumgartner_supercurrent_2022,legg2022superconducting,lotfizadeh2023superconducting,costa2023microscopic,maiani2023nonsinusoidal,hess2023josephson,hou2023ubiquitous} or flux~\cite{kononov2020one,souto2022josephson,gupta2023gate,ciaccia2023gate,valentini2023radio,greco2023josephson,legg2023parity,cuozzo2023microwave} to break time-reversal symmetry.
However, it is also possible for materials to break the relevant symmetries spontaneously at the microscopic level, giving rise to an \textit{intrinsic} SDE~\cite{ZeroFieldDiode,scammell_theory_2022,wu_field-free_2022,zhang2024angle,PhysRevLett.132.046003,diez2023symmetry,hu2023josephson}.
Such an intrinsic SDE could enable field-free non-reciprocal circuit elements~\cite{khabipov2022superconducting,frattini_3-wave_2017,frattini_optimizing_2018,sivak_kerr-free_2019,miano_frequency-tunable_2022,PhysRevApplied.21.064029} and serve as a probe for symmetry-breaking in superconducting materials.

In an exciting recent development, superconductivity has been discovered in rhombohedral tetralayer graphene (RTLG)~\cite{han2024,choi2024}. A unique aspect of RTLG is that a superconducting phase can emerge from a spin- and valley-polarized ``quarter-metal" parent state and persists under large in-plane magnetic fields~\cite{han2024}.
These observations point to a possible spin-polarized pairing with a chiral $p+ip$ order parameter, as suggested by theoretical works~\cite{geier2024chiral,chou2024intravalley,yang2024topological,qin2024chiral,parra2025band,dong2025,jahin2024enhanced,wang2024chiral,yoon2025quarter,maymann2025,christos2025}.
Importantly, the valley polarization in the normal state breaks time-reversal symmetry and in-plane inversion (i.e., two-fold rotational symmetry) \cite{scammell_theory_2022}, making RTLG a promising platform for non-reciprocal superconducting phenomena. Indeed, an observation of an intrinsic SDE in RTLG has recently been reported~\cite{han2024}. However, a proper and quantitative understanding of the diode effect beyond symmetry arguments requires an explicit microscopic calculation.

\begin{figure}[!t]
    \centering
    \includegraphics[width=\linewidth]{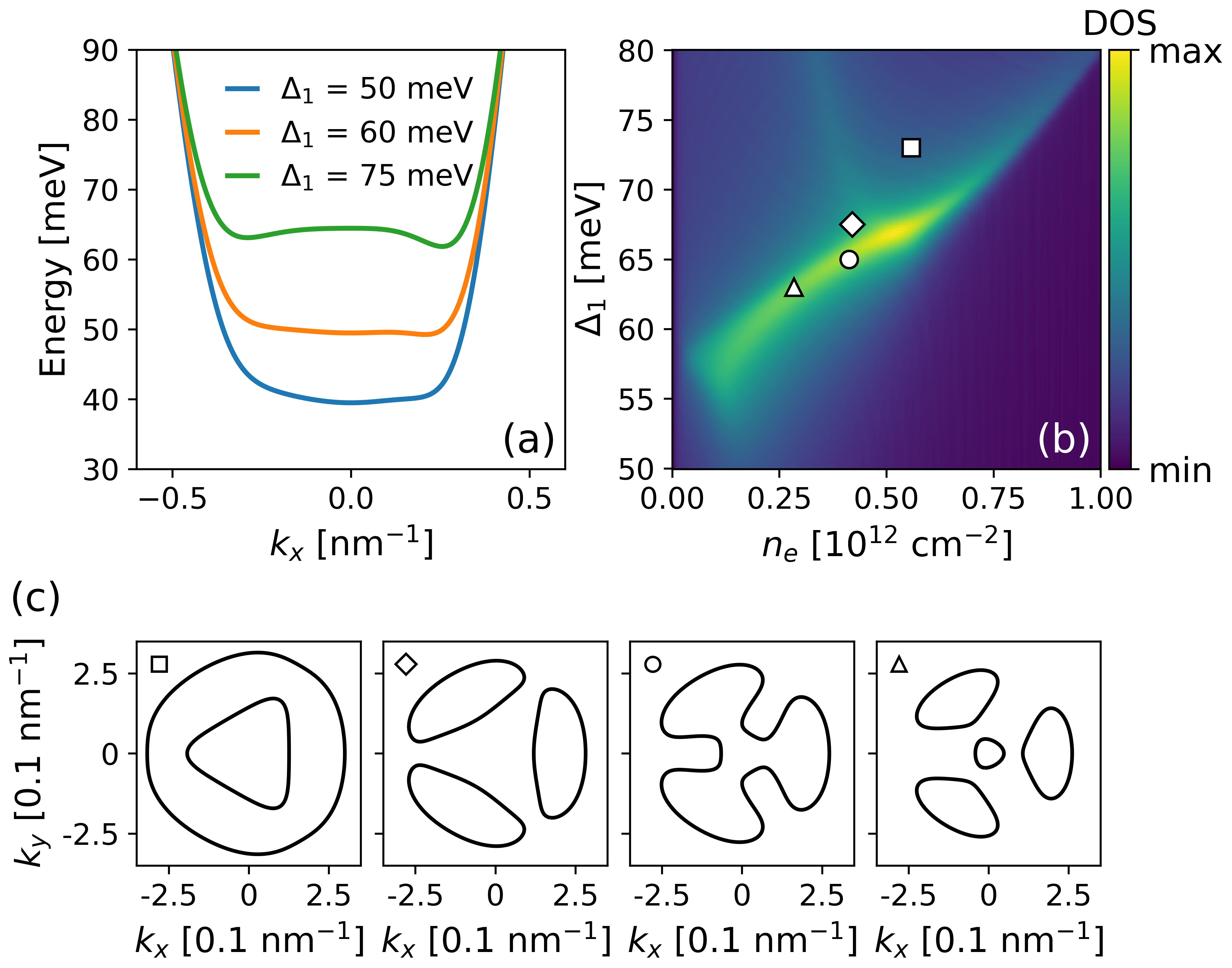}
    \caption{
    (a) Dispersion of RTLG along the $k_x$-direction for different displacement fields, $\Delta_1$, showing a band flattening around $k_{x}=0$. (b) Density of states (DOS) versus $\Delta_1$ and the carrier density, $n_e$ (relative to charge neutrality), showing a significant enhancement due to the band flattening.
    (c) Fermi surfaces for ($\Delta_1,n_e$), from left to right, set to $(72\,\text{meV},0.574)$, $(67.5\,\text{meV},0.386)$, $(65\,\text{meV},0.414)$, $(63\,\text{meV},0.254)$. 
    }
    \label{fig:1}
\end{figure}

Here, we fill this gap by theoretically demonstrating and studying the emergence of nonreciprocal superconductivity and an 
intrinsic SDE in RTLG. Our results are based on a fully self-consistent framework for determining the superconducting order parameter from a Kohn-Luttinger mechanism to superconductivity and show 
that large diode efficiencies are achievable and highly tunable by a displacement field. Moreover, we find that both the diodicity and the critical current in general show a characteristic angular dependence with enhanced lobes, which depend on the Fermi surface topology of the normal state. 
We further show that the same intrinsic non-reciprocity produces an asymmetric current–voltage characteristic in a metal–superconductor junction. 
Taken together, our findings suggest that measurements of supercurrent and conductance asymmetries could provide 
insights into the type of Fermi surface topology from which superconductivity arises.

\textit{Model.} To describe spin- and valley-polarized RTLG, we use an $8$-band $\mathbf{k}\cdot \mathbf{p}$ model for the electronic states near the $K$ or $K'$ valley (labeled by $\tau=\pm$).
The Bloch Hamiltonian, $h_0$, capturing the non-interacting bands, written in the 
basis $(A_{1},B_{1},A_{2},B_{2},A_{3},B_{3},A_{4},B_{4})$, where $A_{\ell}$ and $B_{\ell}$ denote the sublattices of the $\ell^{\text{th}}$ layer, is given by~\cite{Model_Koshino,PhysRevB.82.035409,PhysRevB.107.104502}, 
\begin{align}
\label{Eq1}
&h_{0} = \\
&\begin{pmatrix}
-\Delta_1+\delta & v_0\pi^\dagger & v_4\pi^\dagger & v_3\pi & 0 & \frac{\gamma_2}{2} & 0 & 0 \\
v_0\pi &  -\Delta_1 & \gamma_1 & v_4\pi^\dagger & 0 & 0 & 0 & 0 \\
v_4\pi & \gamma_1 & -\frac{\Delta_1}{3} & v_0\pi^\dagger & v_4\pi^\dagger & v_3\pi & 0 & \frac{\gamma_2}{2} \\
v_3\pi^\dagger & v_4\pi & v_0\pi & -\frac{\Delta_1}{3} & \gamma_1 & v_4\pi^\dagger & 0 & 0 \\
0 & 0 & v_4\pi & \gamma_1 &  \frac{\Delta_1}{3} & v_0\pi^\dagger & v_4\pi^\dagger & v_3\pi \\
\frac{\gamma_2}{2} & 0 & v_3\pi^\dagger & v_4\pi & v_0\pi & \frac{\Delta_1}{3} & \gamma_1 & v_4\pi^\dagger \\
0 & 0 & 0 & 0 & v_4\pi & \gamma_1 & \Delta_1 & v_0\pi^\dagger \\
0 & 0 & \frac{\gamma_2}{2} & 0 & v_3\pi^\dagger & v_4\pi & v_0\pi & \Delta_1+\delta 
\end{pmatrix}.\nonumber
\end{align}
Here $\pi = \tau k_{x} + i k_{y}$ and the momentum $\boldsymbol{k}=(k_{x}, k_{y})$ is measured relative to $K$ or $K'$ valley. 
The $k_{x}$-direction correspond to the $\Gamma$-$K$ direction of the full momentum-space Hamiltonian, while the $k_{y}$-direction corresponds to the $\Gamma$-$M$ direction.
The parameters $v_{j}=\sqrt{3}a\gamma_{j}/2$ are given in terms of the graphene lattice constant, $a=2.46\si{\angstrom}$, and the hoppings, $\gamma_{j}$. Moreover, $\delta$ is a sublattice potential and $\Delta_1$ is an external displacement field. For our simulations, we use $(\gamma_0,\gamma_1,\gamma_2,\gamma_3,\gamma_4,\delta)=(3.1,0.38,-0.015,-0.29,-0.141,-0.0105)\,\text{eV}$~\cite{zhou2021half}.

Diagonalizing $h_0$ in Eq.\,\eqref{Eq1} shows that the low-energy conduction and valence bands in RTLG touch near the $K$ or $K'$ valley and originate 
primarily from the weakly-coupled $A_1$ and $B_4$ sublattices. The application of a displacement field, $\Delta_1$, opens an energy gap and flattens these lowest bands, leading to a significant enhancement of the DOS, see Fig.~\ref{fig:1}(a) and (b). Furthermore, a finite $\Delta_1$ reverses the band curvature near $\boldsymbol{k}=0$. As a result, the Fermi surface can display multiple connected components depending on $\Delta_1$ and the electron density $n_e$, see Fig.~\ref{fig:1}(c).

The aforementioned flattening of the low-energy bands in RTLG enhances the importance of interaction effects. 
In the recent observation of superconductivity from a spin- and valley-polarized ``quarter-metal", the chemical potential was tuned to the bottom of the lowest conduction band~\cite{han2024}.
This motivates us to introduce the following effective Hamiltonian for the interacting system, 
\begin{equation}
\label{Eq2}
\begin{split}
H_{\text{eff}} &= \sum_{\boldsymbol{k}}    
\varepsilon_{\boldsymbol{k}}
\,
c^\dagger_{\boldsymbol{k}}c^\pdagger_{\boldsymbol{k}}
\\
&
+
\frac{1}{2\Omega} 
\sum_{\boldsymbol{k},\boldsymbol{k}',\boldsymbol{q}}
\tilde{V}_{\boldsymbol{q}}\,
F_{\boldsymbol{k},\boldsymbol{k}+\boldsymbol{q}}\,
F_{\boldsymbol{k}',\boldsymbol{k}'-\boldsymbol{q}}\; 
c_{\boldsymbol{k}}^\dagger
c^\pdagger_{\boldsymbol{k}+\boldsymbol{q}}
c_{\boldsymbol{k}'}^\dagger
c^\pdagger_{\boldsymbol{k}'-\boldsymbol{q}}
\end{split}
\end{equation}
The first term describes the dispersion, $\varepsilon_{\boldsymbol{k}}$, of the lowest conduction band obtained from $h_0$ with $c_{\boldsymbol{k}}$ denoting the electron annihilation operator of the active spin and valley flavor in that band. 
\begin{figure}[!t]
    \centering
    \includegraphics[width=\linewidth]{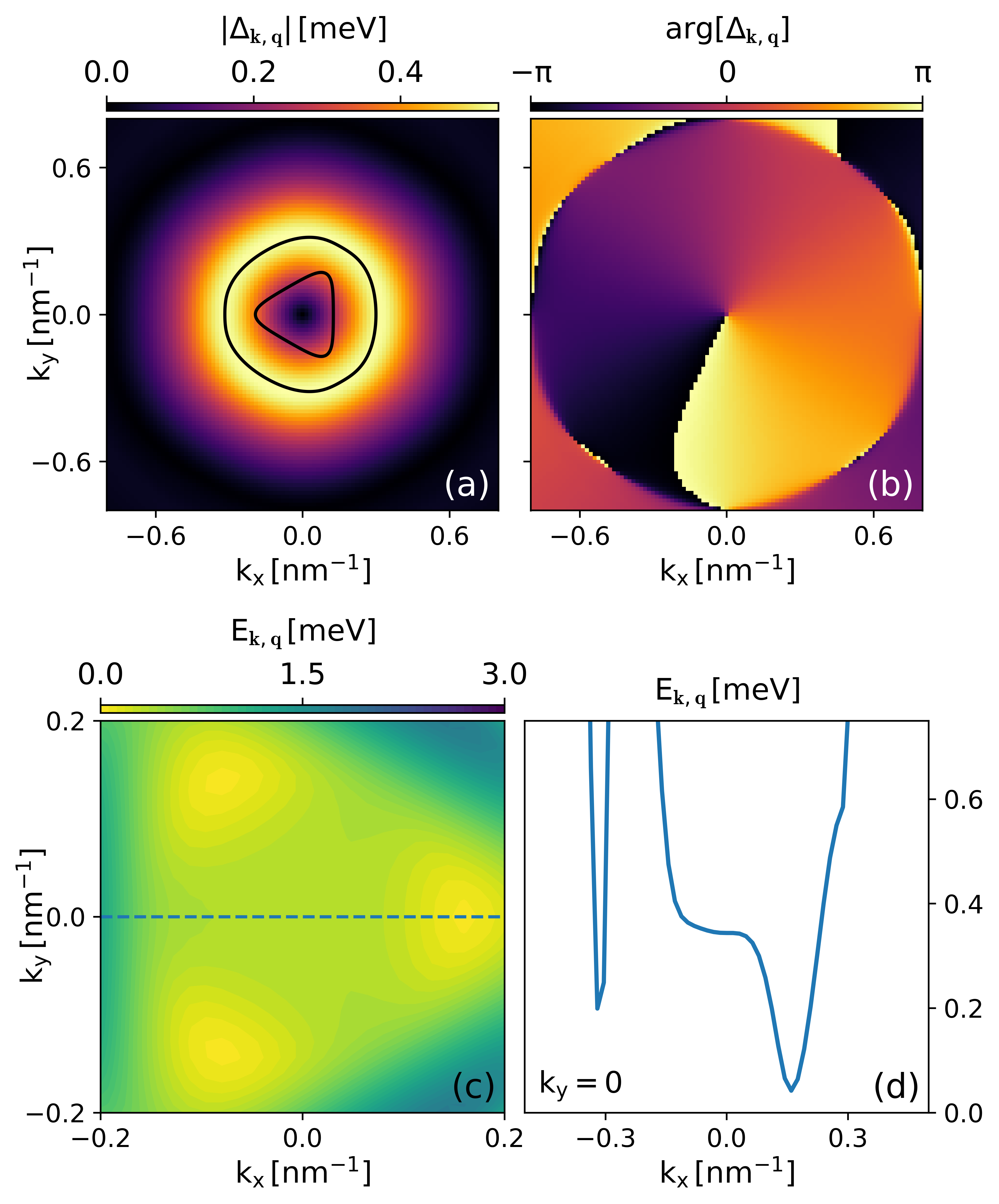}
    \caption{
    (a) Magnitude of the superconducting order parameter, $\Delta_{\boldsymbol{k},\boldsymbol{q}}$, for the Fermi surface shown as an inset in the zero-temperature limit.
    (b) Phase of the superconducting order parameter, showing a $p+ip$ winding.  Here, we have chosen a gauge such that the wavefunction of the lowest conduction band has real-valued inner product with the wavefunction at $\boldsymbol{k}=0$, i.e., $\braket{\psi_{\boldsymbol{k}}}{\psi_{\boldsymbol{0}}}$ is real. (c) Quasiparticle dispersion, $E_{\boldsymbol{k,\boldsymbol{q}}}$, associated with the superconducting order parameter shown in (a) and (b). (d) The quasiparticle dispersion is asymmetric under $k_{x}\rightarrow -k_{x}$. Hence, RTLG realizes a nonreciprocal superconductor. 
    }
    \label{fig:2}
\end{figure}
The second term captures interaction effects by projecting a repulsive Coulomb interaction 
onto the lowest conduction band and accounting for screening effects via a random phase approximation (RPA). This projection introduces the form factors, $F_{\boldsymbol{k},\boldsymbol{k}+\boldsymbol{q}} = \langle \psi_{\boldsymbol{k}}|\psi_{\boldsymbol{k}+\boldsymbol{q}}\rangle$, that describe the overlap of Bloch states in the lowest conduction band. The screened Coulomb potential is
\begin{equation}
\label{Eq3}
\tilde{V}_{\boldsymbol{q}}
=
\frac{V_{\boldsymbol{q}}}{1-\Pi_{\mathbf{q}}V_{\boldsymbol{q}}},
\end{equation}
where $V_{\boldsymbol{q}} = \frac{e^2}{2\epsilon} \frac{\tanh(|\boldsymbol{q}|d)}{|\boldsymbol{q}|}$ denotes the gate-screened Coulomb potential with $d$ being the distance to the metallic gates and $\epsilon$ the dielectric permittivity of the hexagonal boron nitride encapsulation. We take $d=20\,\text{nm}$ and $\epsilon = 6\epsilon_0$ with $\epsilon_0$ the vacuum permittivity. The polarization function, $\Pi_{\boldsymbol{q}}$, entering Eq.\,\eqref{Eq3} is given by
$
\Pi_{\boldsymbol{q}} = \frac{1}{\Omega} \sum_{\boldsymbol{k}}   
\frac{
n_{F}(\varepsilon_{\boldsymbol{k}+\boldsymbol{q}})
-
n_{F}(\varepsilon_{\boldsymbol{k}})
}{
\varepsilon_{\boldsymbol{k}+\boldsymbol{q}}
-
\varepsilon_{\boldsymbol{k}}
}
|F_{\boldsymbol{k},\boldsymbol{k}+\boldsymbol{q}}|^2
$,
where $n_{F}$ is the Fermi-Dirac distribution and $\Omega$ the area of the system. 
We will now use this effective Hamiltonian to explain the emergence of superconductivity, study the resulting order parameter, and demonstrate its non-reciprocal signatures, including a sizable intrinsic SDE.

\textit{Nonreciprocal superconductivity.} 
We begin by discussing nonreciprocal superconductivity. Nonreciprocal superconductors are a recently proposed class of superconductors in which the breaking of time-reversal and inversion symmetry leads to an asymmetric quasiparticle dispersion~\cite{davydovageier2024}. Here, we will explicitly show that in RTLG this asymmetric quasiparticle dispersion arises from the asymmetry of the normal-state band structure, $\varepsilon_{\boldsymbol{k}} \neq \varepsilon_{-\boldsymbol{k}}$, and discuss how a finite Cooper-pair momentum (relative to $K$ or $K'$)~\cite{yang2024topological,qin2024chiral} can influence the the symmetry properties of quasiparticle dispersion.

To describe the superconductivity in RTLG, we consider, inspired by earlier works~\cite{geier2024chiral,chou2024intravalley,yang2024topological,qin2024chiral,parra2025band,dong2025}, a Kohn-Luttinger mechanisms~\cite{kohn1965new,chubukov1993kohn,PhysRevB.107.104502,PhysRevLett.127.247001,PhysRevB.110.035143}. Specifically, while the bare Coulomb potential, $V_{\boldsymbol{q}}$, is nominally repulsive, 
the screening effects described by Eq.\,\eqref{Eq3} can lead to an ``overscreening" regime, where the effective interaction becomes attractive in some channels. This effective attraction can be verified by transforming $\tilde{V}_{\mathbf{q}}$ to real space.

We will assume that this effective attraction can lead to pairing between momentum states $\boldsymbol{k}+\frac{\boldsymbol{q}}{2}$ and $-\boldsymbol{k}+\frac{\boldsymbol{q}}{2}$ where $\boldsymbol{q}$ is the Cooper-pair momentum. While we retain the possibility of a finite $\boldsymbol{q}$, we emphasize that it is not a necessary ingredient for the nonreciprocal superconductivity and SDE. We then perform a mean-field decoupling of Eq.\,\eqref{Eq2} in the Cooper channel. The superconducting order parameter is determined self-consistently as, 
\begin{equation}
\begin{split}
\Delta_{\boldsymbol{k},\boldsymbol{q}}
=
&-\frac{1}{\Omega}
\sum_{\boldsymbol{k}'}
g_{\boldsymbol{k},\boldsymbol{k}',\boldsymbol{q}} \, 
\frac{
\Delta_{\boldsymbol{k}',\boldsymbol{q}}
}{
2 \tilde{E}_{\boldsymbol{k}',\boldsymbol{q}}
}
\\
\times
&
\frac{1}{2}
\left\{
\tanh
\left(
\frac{\beta E_{\boldsymbol{k}',\boldsymbol{q}}}{2}
\right)
+
\tanh
\left(
\frac{\beta E_{-\boldsymbol{k}',\boldsymbol{q}}}{2}
\right)
\right\}.
\end{split}
\label{Eq5}
\end{equation}
where 
$
g_{\boldsymbol{k},\boldsymbol{k}',\boldsymbol{q}}
=
\tilde{V}_{\boldsymbol{k}'-\boldsymbol{k}}
F_{\boldsymbol{k}+\frac{\boldsymbol{q}}{2},\boldsymbol{k}'+\frac{\boldsymbol{q}}{2}}
F_{-\boldsymbol{k}+\frac{\boldsymbol{q}}{2},-\boldsymbol{k}'+\frac{\boldsymbol{q}}{2}}
$ is the pairing interaction, $1/\beta=k_{B}T$, 
and the quasiparticle dispersion is given by, 
\begin{equation}
\label{Eq6}
E_{\boldsymbol{k},\boldsymbol{q}}
=
\xi_{\boldsymbol{k},\boldsymbol{q},-}
+
\sqrt{
(\xi_{\boldsymbol{k},\boldsymbol{q},+})^{2}
+
|
\Delta_{\boldsymbol{k},\boldsymbol{q}}
|^{2}
}
\end{equation}
Here, we have defined the symmetric and antisymmetric components of the normal-state dispersion as
$
\xi_{\boldsymbol{k},\boldsymbol{q},\pm}
=
(
\varepsilon_{\boldsymbol{k}+\frac{\boldsymbol{q}}{2}}
\pm
\varepsilon_{-\boldsymbol{k}+\frac{\boldsymbol{q}}{2}}
)/2
$. Moreover, we have introduced the reduced quasiparticle dispersion, $\tilde{E}_{\boldsymbol{k}',\boldsymbol{Q}}=E_{\boldsymbol{k},\boldsymbol{q}}-\xi_{\boldsymbol{k},\boldsymbol{q},-}$.

\begin{figure}[!t]
    \centering
    \includegraphics[width=\linewidth]{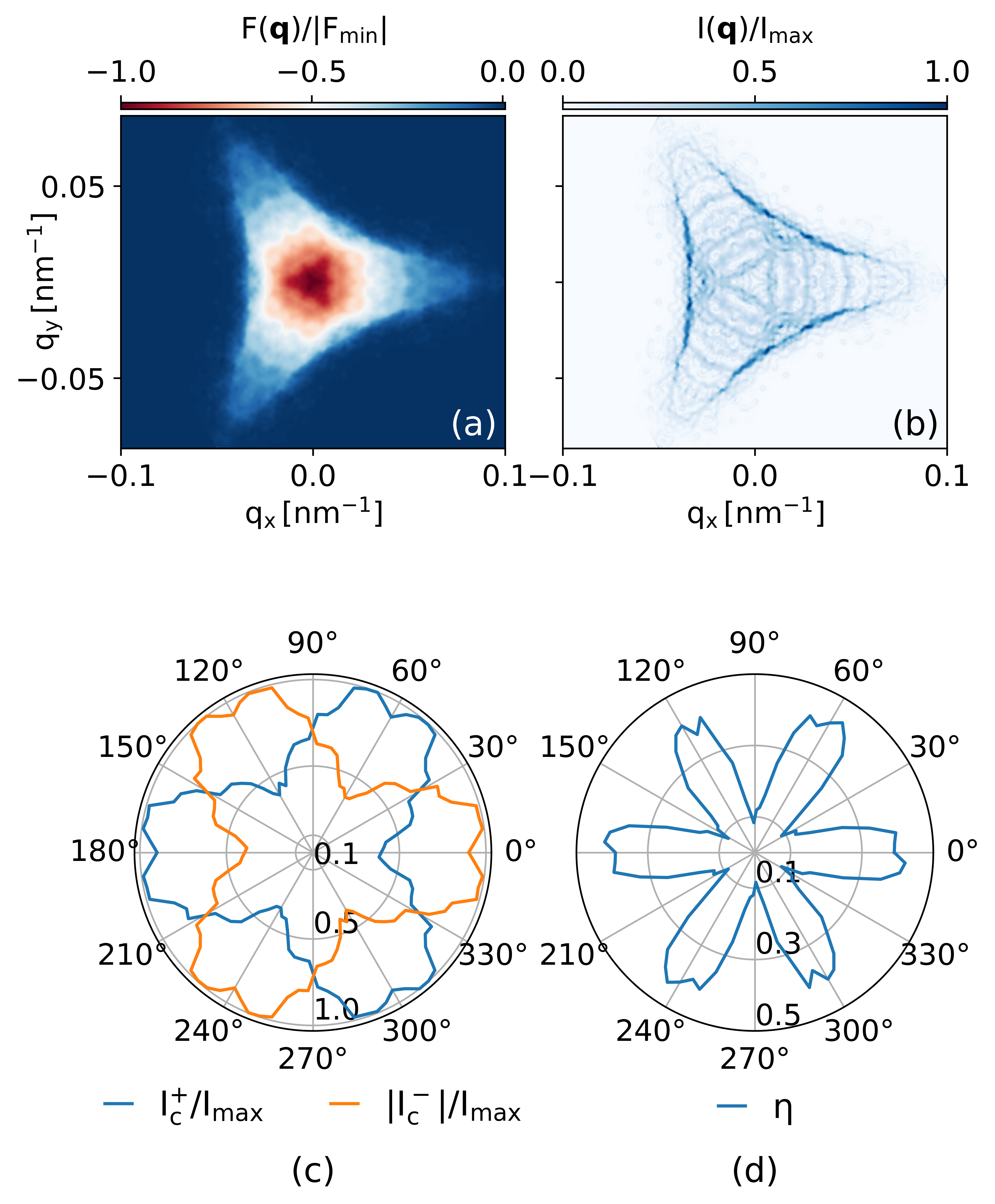}
    \caption{
    (a) Free energy, $F$, versus $\boldsymbol{q}=(q_x,q_y)$ for the annulus Fermi surface shown in Fig.\,\ref{fig:2}. 
    Here, $F_{\text{min}}\equiv\text{min}_{\boldsymbol{q}}\{F(\boldsymbol{q})\}.$
    (b) Corresponding supercurrent magnitude, $I$, versus $\boldsymbol{q}$. Here, $I_{\text{max}}\equiv\text{max}_{\boldsymbol{q}}\{I(\boldsymbol{q})\}$. 
    (c) Critical currents in the forward direction ($I^{+}_{c}$) and reverse direction ($I^{-}_{c}$) for current direction given by an angle $\theta$ relative to the $x$-direction. 
    (d) Diode efficiency for a current direction given by an angle $\theta$.
    }
    \label{fig:3}
\end{figure}

Next, we solve for $\Delta_{\boldsymbol{k},\boldsymbol{q}}$ self-consistently, imposing a fixed electron density $n_{e} = \sum_{\boldsymbol{k}} \langle c^\dagger_{\boldsymbol{k}}c_{\boldsymbol{k}} \rangle / \Omega$ and varying $\boldsymbol{q}$. The realized superconducting order parameter, $\Delta_{\boldsymbol{k},\boldsymbol{q}}$, is determined by minimizing the free energy, 
\begin{align}
\label{Eq7}
F 
&= 
-
\sum_{\boldsymbol{k}}
\left\{
\frac{1}{\beta}
\ln
(
1 + e^{-\beta E_{\boldsymbol{k},\boldsymbol{q}}}
)
-
\frac{1}{2}
(
\xi_{\boldsymbol{k},\boldsymbol{q},+}
-
\tilde{E}_{\boldsymbol{k},\boldsymbol{q}}
)
\right.
\\
&\qquad-
\left.
\frac{
\abs{\Delta_{\boldsymbol{k},\boldsymbol{q}}}^2
}{
8 \tilde{E}_{\boldsymbol{k},\boldsymbol{q}}
}
\left[
\tanh
\left(
\frac{\beta E_{\boldsymbol{k},\boldsymbol{q}}}{2}
\right)
+
\tanh
\left(
\frac{\beta E_{-\boldsymbol{k},\boldsymbol{q}}}{2}
\right)
\right]
\right\}
\nonumber
\end{align}

An example result of our self-consistency calculation is shown in Fig.\,\ref{fig:2}(a) and (b) for a Fermi surface with an annulus topology. We see that the superconducting order parameter exhibits a 
$p+ip$ phase winding. 
The Cooper-pair momentum is found to be small but finite (${q}_{0}=-\SI{0.004}{\nano\meter^{-1}}$), oriented along the $q_x$-direction (i.e., the $\Gamma$-$K$ direction) or the two directions related by the three-fold out-of-plane rotational symmetry, $C_{3}$, of the system. 
Using the self-consistent solution, we then compute the quasiparticle dispersion, $E_{\boldsymbol{k},\boldsymbol{q}}$, see Fig.\,\ref{fig:2}(c).
The quasiparticle dispersion is in general found to be asymmetric with respect to $\boldsymbol{k}\rightarrow-\boldsymbol{k}$. Hence, RTLG realizes nonreciprocal superconductivity. Writing $\boldsymbol{k}=(k\cos\theta,k\sin\theta)$, we see that the asymmetry is most pronounced along $\theta\in\{0^\circ,60^\circ,120^\circ\}$. It is instructive to discuss these results from a symmetry perspective.

First, assume $\boldsymbol{q}=0$. On top of the aforementioned $C_{3}$ symmetry, the normal state spectrum is invariant under reflection $\sigma_v$ at the mirror planes along $\theta\in\{0^\circ,60^\circ,120^\circ\}$.

Consider 
a $\boldsymbol{k}$ pointing along one of the perpendicular directions 
$\theta\in\{90^\circ,150^\circ,30^\circ\}$. Then, $\sigma_v\boldsymbol{k}=-\boldsymbol{k}$ and hence $\varepsilon_{-\boldsymbol{k}}=\varepsilon_{\sigma_v\boldsymbol{k}}=\varepsilon_{\boldsymbol{k}}$. As a result, 
$\xi_{\boldsymbol{k},\boldsymbol{q},-}=0$ and $E_{\boldsymbol{k},\boldsymbol{q}}=E_{-\boldsymbol{k},\boldsymbol{q}}$ (since $|\Delta_{\boldsymbol{k},\boldsymbol{q}}|=|\Delta_{-\boldsymbol{k},\boldsymbol{q}}|$). Thus, the quasiparticle spectrum is symmetric along $\theta\in\{90^\circ,150^\circ,30^\circ\}$, but asymmetric otherwise. 

Now assume $\boldsymbol{q}\neq0$. In this case, a vanishing asymmetry of the quasiparticle spectrum requires both 
$\sigma_v\boldsymbol{k}=-\boldsymbol{k}$ and $\sigma_v\boldsymbol{q}=\boldsymbol{q}$. For  
$\boldsymbol{q} = (q_x,0)^T$, i.e., pointing along $\theta=0^\circ$, the two conditions can only be satisfied for momenta $\boldsymbol{k}$ with $\theta=90^\circ$. 
Consequently, the quasiparticle dispersion can be strictly symmetric only along $\theta=90^\circ$; in all other directions, it remains asymmetric. Note, however, that as a result of the numerically small value of $q_x$, the dispersion shown in Fig.\,\ref{fig:2}(c) still looks approximately symmetric along $\theta=90^\circ, 150^\circ$, and $30^\circ$.

One experimental signature of the asymmetric quasiparticle dispersion is that a transparent normal-metal–superconductor junction can exhibit a current-voltage ($I$-$V$) characteristic that is not invariant under $V\rightarrow-V$~\cite{davydovageier2024}. Our results suggest the following probe of the nonreciprocal superconductivity in RTLG: connect two normal leads to the RTLG superconductor, one along $\theta=0^\circ$ ($\Gamma$-$K$ direction) and one along $\theta=90^\circ$ ($\Gamma$-$M$ direction). Our results imply that the lead along $\theta=0^\circ$ would show an asymmytric $I$-$V$ curve, while the $I$-$V$ curve along $\theta=90^\circ$ would remain symmetric. 

\textit{Intrinsic SDE.} 
Next, we address the intrinsic SDE. 
For this purpose, we self-consistently solve for the superconducting order parameter for a broad range of Cooper-pair momenta, $\boldsymbol{q}$, and use the obtained $\Delta_{\boldsymbol{k},\boldsymbol{q}}$ to evaluate the free energy. The results of the free energy are shown in Fig.\,\ref{fig:3}(a). The current in the ground state that minimizes $F(\mathbf{q})$ is always zero. However, applying a finite current-bias, produces a state with a finite Cooper-pair momentum different from the ground state value, $\boldsymbol{q}\neq\boldsymbol{q}_0$. The current in this state is $I=2e|\partial_{\boldsymbol{q}}F(\boldsymbol{q})|$\,\cite{daido2022intrinsic}, see Fig.\,\ref{fig:3}(b).

We then fix a direction $\theta$, so that $\boldsymbol{q}=(q\cos\theta,q\sin\theta)$ and determine the 
 critical currents in the forward direction, 
$I^{+}_{c}(\theta)
=
I_{c}(\theta)
$
and 
 in the reversed direction,
$I^{-}_{c}(\theta)
=
I_{c}(\theta+\pi)
$.
The diode efficiency is defined as,
\begin{equation}
\eta = \frac{|I^{+}_{c}-I^{-}_{c}|}{I^{+}_{c}+I^{-}_{c}}.   
\end{equation}
Here, $\eta$ is non-negative. However, we remark that the difference in critical currents, $\Delta I_{c}=I^{+}_{c}-I^{-}_{c}$,
can exhibit sign changes when transitioning through $\eta=0$, see \cite{Supplemental}.

\begin{figure}[!t]
    \centering
    \includegraphics[width=\linewidth]{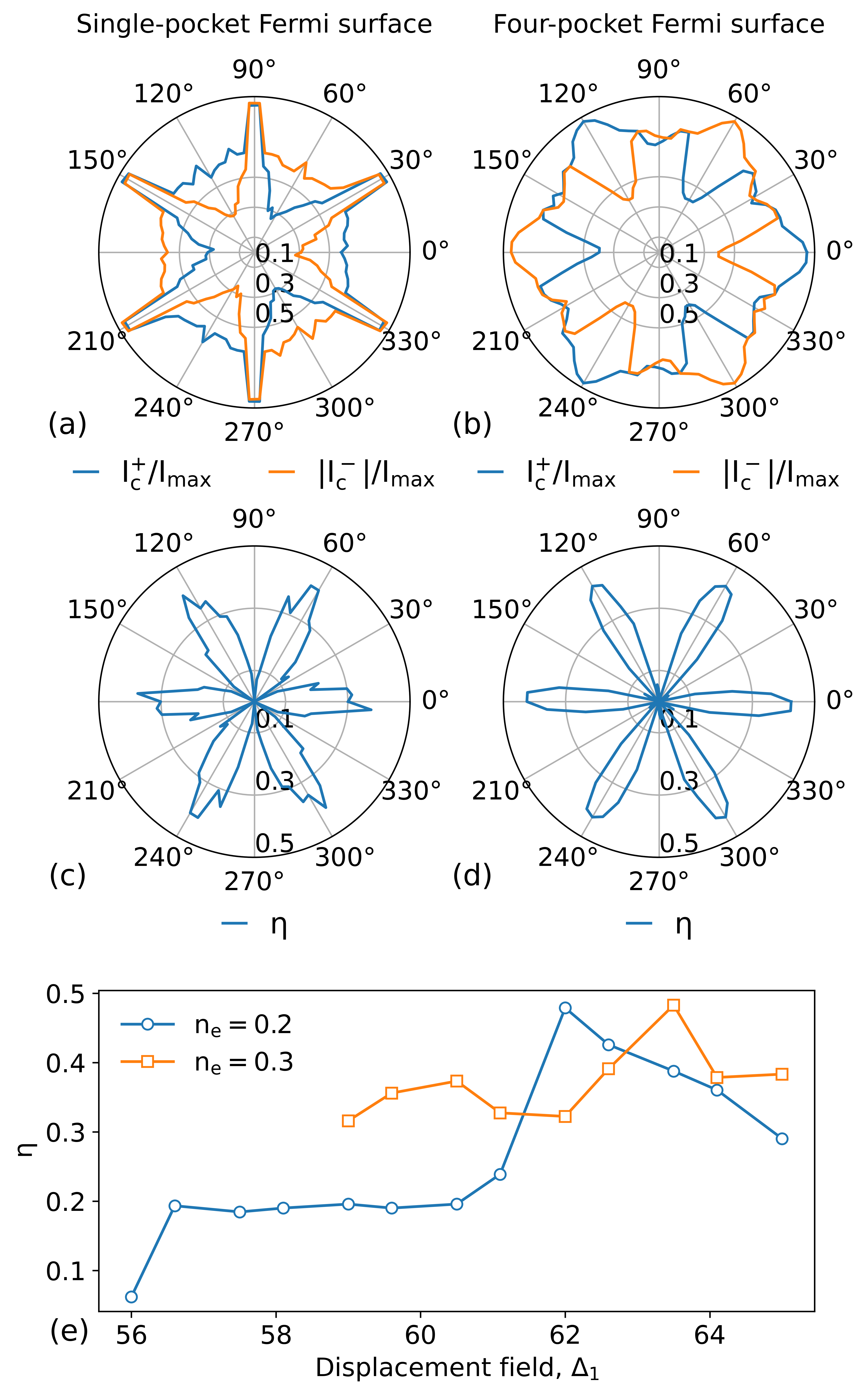}
    \caption{
    (a) Angular dependence of the diode efficiency, $\eta$, for the single-pocket Fermi surface shown in 
    Fig.\,\ref{fig:1}(c). 
    (b) Same as (a) but for the four-pocket Fermi surface shown in 
    Fig.\,\ref{fig:1}(c). 
    (c) Diode efficiency, $\eta$, versus the displacement field along the $q_x$ direction for different electron densities, $n_e$. 
    }
    \label{fig:4}
\end{figure}

We begin by evaluating the angular dependence of the critical currents and diode efficiencies for fixed $\Delta_1$ and $n_e$. Our results are shown in Fig.\,\ref{fig:3}(c) and (d) for a Fermi surface with annulus topology. We find that, in general, the diode efficiency takes on finite values. This verifies the emergence of an intrinsic SDE from a fully self-consistent calculation. Furthermore, we find that $\eta$
exhibits a characteristic multi-lobe structure. Specifically, we see that the efficiency tends to be maximized along the directions $\theta\in\{0^\circ,60^\circ,120^\circ\}$, reaching values of $\sim30\%$ for the annulus Fermi surface. In addition, as a result of the reflection symmetry $\sigma_v$, $\eta$ vanishes along the directions perpendicular to one of the three mirror planes \cite{PhysRevB.110.024503}, i.e., for $\theta\in\{30^\circ,90^\circ,150^\circ\}$.

It is interesting to compare these results to other Fermi surface topologies. In Fig.\,\ref{fig:4}(a–d), we present our findings for both a single-pocket and a four-pocket Fermi surface. As shown in Fig.\,\ref{fig:4}(a) and (b), the angular dependence of the forward and reverse critical currents, $I_c^{\pm}$, differs markedly between the two cases. The single-pocket Fermi surface exhibits sharp peaks in $I_c^{\pm}$ that are absent for the four-pocket topology. These differences are also reflected in the diode efficiencies shown in Fig.\,\ref{fig:4}(c) and (d), which are more sharply peaked in the single-pocket case. Compared to the previous annular Fermi surface discussed in Fig.\,\ref{fig:3}, the lobes for both the single-pocket and four-pocket cases also appear narrower. Altogether, these results indicate that the Fermi surface topology strongly influences the \textit{angular dependence} of both the critical currents and the diode efficiency. In particular, the angular profile of the diode efficiency and the critical currents could provide insights into the type of Fermi surface topology from which the superconducting state arises. More detailed plots of the free energies and supercurrents for single- and four-pocket Fermi surfaces are shown in the Supplemental Material \cite{Supplemental}.

Finally, we also briefly comment on the dependence of the diode efficiency on $\Delta_1$. Our results of the diode efficiencies along the $x$-direction versus $\Delta_1$ for different densities, $n_e$, are shown in Fig.\,\ref{fig:4}(c). We find that the diode efficiency tends to increase with increasing $\Delta_1$. However, we also observe that when increasing $\Delta_1$ above a threshold, the superconducting gap closes and the system enters a metallic phase.

\textit{Conclusion.}
We have theoretically studied the intrinsic SDE as a novel probe of the candidate chiral superconducting state in RTLG. In particular, we have argued that the angular dependence of the critical current asymmetry and of the SDE in this system can provide a route to determine the Fermi surface topology from which superconductivity arises. We note that these angular dependencies can be probe experimentally using sunflower-shaped samples \cite{zhang2024angle,JiaNew}. We hope that our work will contribute to understanding novel superconducting states in RTLG and to the development of novel techniques for their detection\,\cite{sedov2025}.

\textit{Note added.} We have recently become aware of a related work~\cite{yoon2025quarter}. 

\textit{Acknowledgement.} We acknowledge helpful discussions with Marco Valentini and Sayan Banerjee.

\begin{widetext}
\newpage

\begin{center}
\large{\bf Supplemental Material to `Intrinsic superconducting diode effect and nonreciprocal superconductivity in rhombohedral graphene multilayers' \\}
\end{center}
\begin{center}Yinqi Chen$^{1}$, Mathias S. Scheurer$^{2}$, Constantin Schrade$^{1}$
\\
{\it $^{1}$Hearne Institute of Theoretical Physics, Department of Physics \& Astronomy, Louisiana State University, Baton Rouge LA 70803, USA}
\\
{\it $^{2}$Institute for Theoretical Physics III, University of Stuttgart, 70550 Stuttgart, Germany}
\end{center}
In the Supplemental Material, we provide details on the angular dependence of the diode efficiency and the
the mean-field approach for describing the superconducting RTLG system.

\setcounter{secnumdepth}{2}          
\renewcommand{\thesection}{\arabic{section}}        
\renewcommand{\thesubsection}{\thesection.\arabic{subsection}}
\setcounter{section}{0}

\section{Angular-dependence of the diode efficiency}
Here, we provide details on the angular dependence of the diode efficiency for different Fermi surfaces. We also suggest a possible experimental setup to probe the angular dependence of the diodicity.
\\
\\
Specifically, the obtained angular dependence of the diodicity for a single-pocket Fermi surface and a four-pocket Fermi surface (as shown in Fig.\,1(c) in the main text) are given in Fig.\,\ref{fig:1sm} and Fig.\,\ref{fig:2sm}. We also show the signed diode efficiency, $\eta_{signed}$, for the annulus Fermi surface in Fig.\,\ref{fig:3sm}. 
\\
\\
\begin{figure}[!b]
    \centering
    \includegraphics[width=0.5\linewidth]{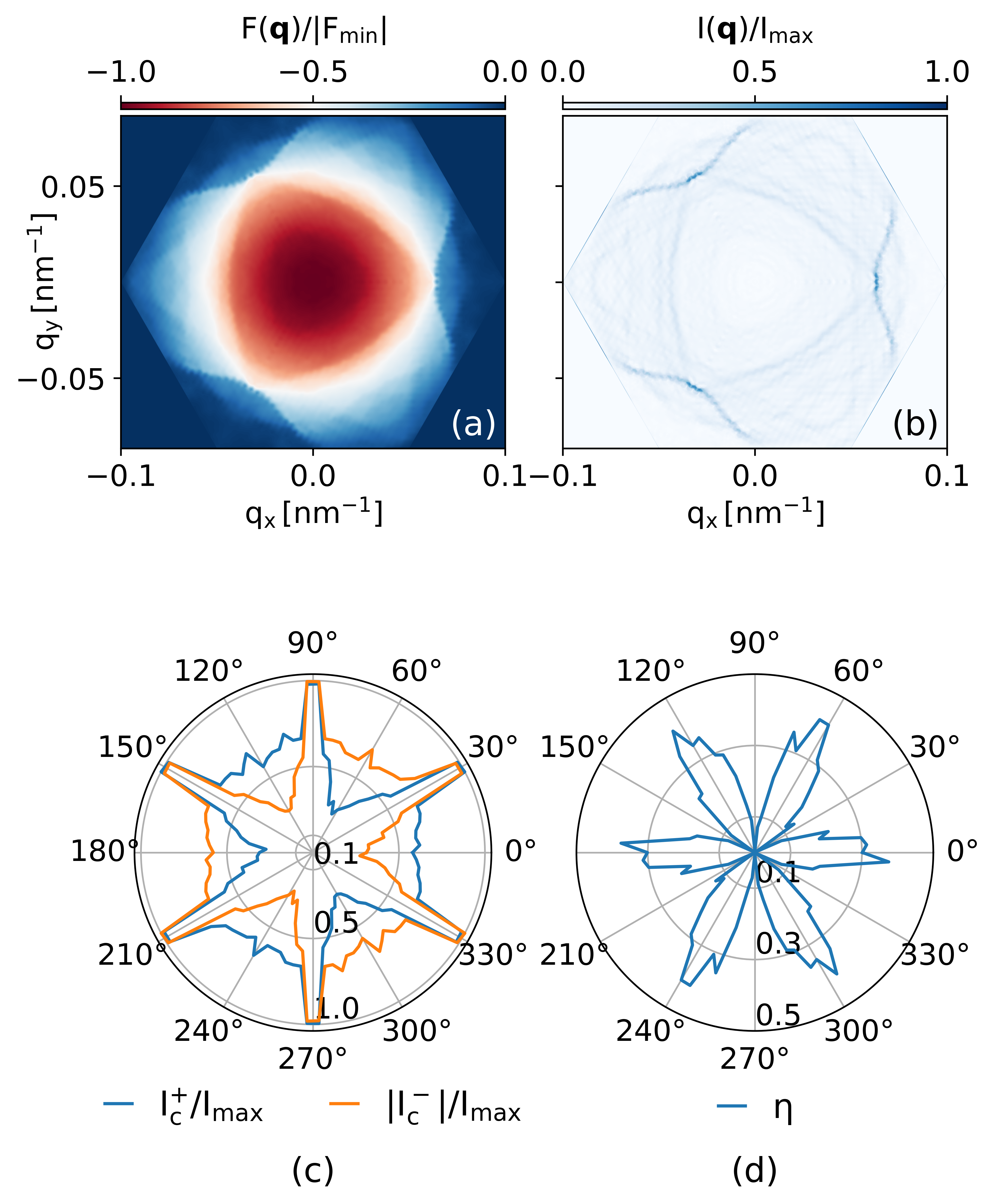}
    \caption{
    Results for a Fermi surface with \textit{one connected components} (as shown in Fig.\,1(c) of the main text). 
    (a) Free energy, $F$, as a function of $\boldsymbol{q}=(q_x,q_y)$.
    Here, $F_{\text{min}}\equiv\text{min}_{\boldsymbol{q}}F(\boldsymbol{q}).$
    (b) Corresponding supercurrent magnitude, $I$, as a function of $\boldsymbol{q}$. Here, $I_{\text{max}}\equiv\text{max}_{\boldsymbol{q}}I(\boldsymbol{q})$. 
    (c) Critical currents in the forward direction ($I_{c,+}$) and reverse direction ($I_{c,-}$) for current direction given by an angle $\theta$ relative to the $x$-direction. 
    (d) Diode efficiency for a current direction given by an angle $\theta$.
    }
    \label{fig:1sm}
\end{figure}
To measure the angular dependence of the diode efficiency, we propose a possible experimental setup, which is schematically shown in Fig.\,\ref{fig:4sm} and inspired by \cite{zhang2024angle}. In the proposed setup, a circular superconducting RTLG system is connected to multiple leads. Applying a current-bias across the different possible directions would allow for a measurement of the critical currents and the diodicity along the different directions.

\begin{figure}[!h]
    \centering
    \includegraphics[width=0.5\linewidth]{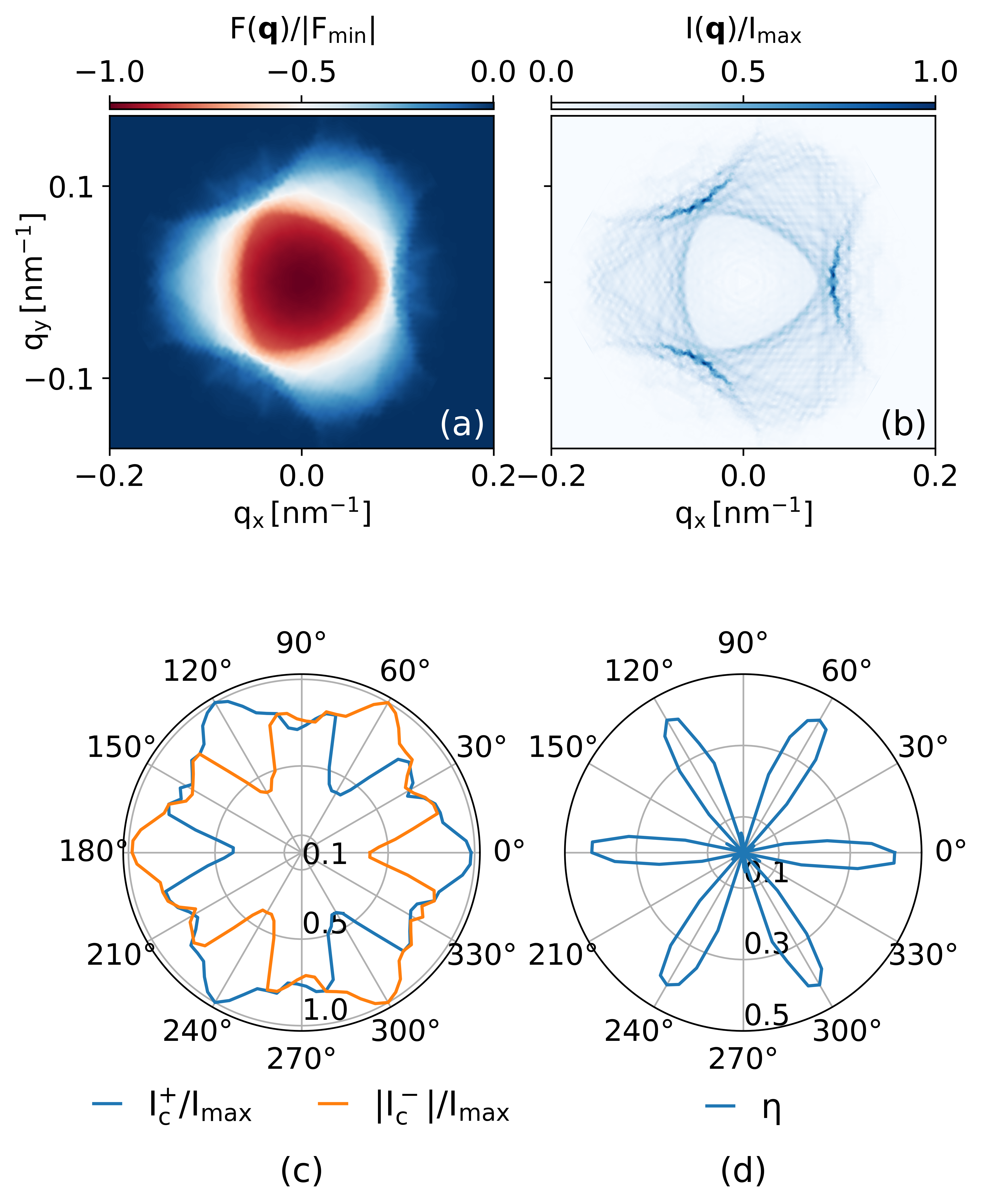}
    \caption{
    Results for a Fermi surface with \textit{four connected components} (as shown in Fig.\,1(c) of the main text). 
    (a) Free energy, $F$, as a function of $\boldsymbol{q}=(q_x,q_y)$.
    Here, $F_{\text{min}}\equiv\text{min}_{\boldsymbol{q}}F(\boldsymbol{q}).$
    (b) Corresponding supercurrent magnitude, $I$, as a function of $\boldsymbol{q}$. Here, $I_{\text{max}}\equiv\text{max}_{\boldsymbol{q}}I(\boldsymbol{q})$. 
    (c) Critical currents in the forward direction ($I_{c,+}$) and reverse direction ($I_{c,-}$) for current direction given by an angle $\theta$ relative to the $x$-direction. 
    (d) Diode efficiency for a current direction given by an angle $\theta$.
    }
    \label{fig:2sm}
\end{figure}

\begin{figure}[!h]
    \centering
    \includegraphics[width=0.48\linewidth]{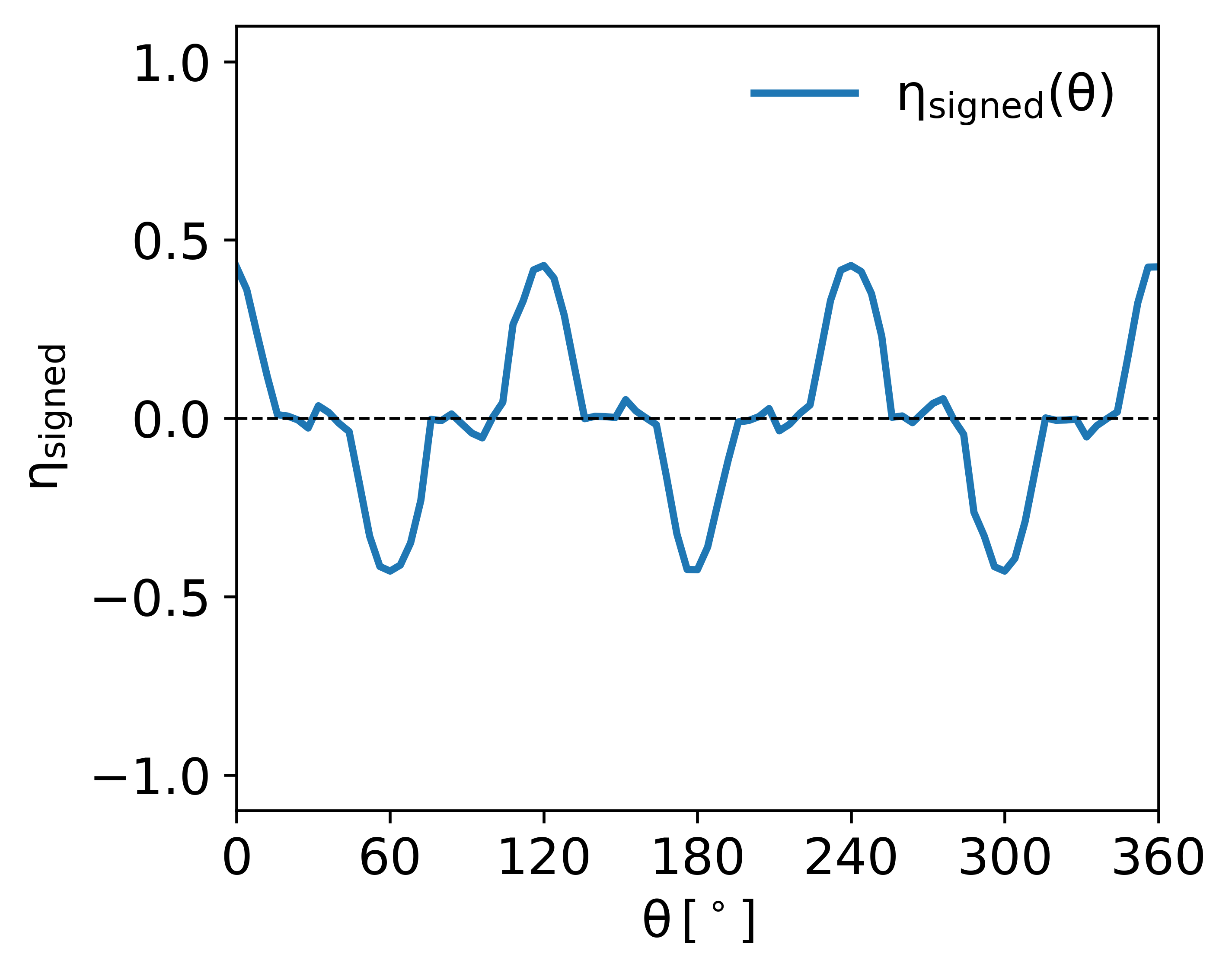}
    \caption{
    Signed diode efficiency, $\eta_{\text{signed}}$ as a function of the angle, $\theta$, for the annulus fermi surface discussed in Fig.\,3 of the main text.
    }
    \label{fig:3sm}
\end{figure}

\begin{figure}[!h]
    \centering
    \includegraphics[width=0.4\linewidth]{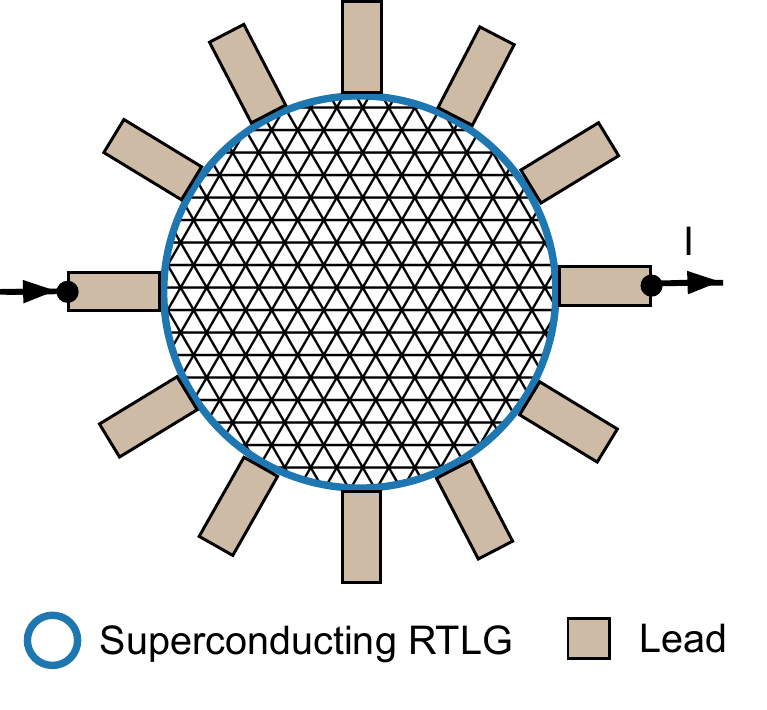}
    \caption{
Possible setup for determining the angular dependence of diode efficiency in RTLG, inspired by \cite{zhang2024angle}. A circular superconducting RTLG sample is connected to multiple leads. Application of a current bias, $I$, along the different directions
enables the extraction of the critical currents and diode efficiencies.
    }
    \label{fig:4sm}
\end{figure}

\newpage

\section{Mean-field decoupling}
Here, we provide a detailed discussion of the mean-field decoupling of the effective interaction Hamiltonian.
\\
\\
We consider only scattering processes of electron pairs with a finite Cooper-pair momentum, $\boldsymbol{q}$. Then the effective Hamiltonian of the interacting systems can be written as,
\begin{equation}
\begin{split}
H_{\text{eff}} &\rightarrow 
\sum_{\boldsymbol{k}}    
\varepsilon_{\boldsymbol{k}}
\,
c^\dagger_{\boldsymbol{k}}c_{\boldsymbol{k}}
\\
&
+
\frac{1}{2\Omega} 
\sum_{\boldsymbol{k},\boldsymbol{k}',\boldsymbol{q}}
g_{\boldsymbol{k},\boldsymbol{k}',\boldsymbol{q}}\,
c_{\boldsymbol{k} + \frac{\boldsymbol{Q}}{2}}^\dagger
c_{-\boldsymbol{k} + \frac{\boldsymbol{Q}}{2}}^\dagger
c_{-\boldsymbol{k'} + \frac{\boldsymbol{Q}}{2}}
c_{\boldsymbol{k'} + \frac{\boldsymbol{Q}}{2}}
\end{split}
\end{equation}
with the pairing interaction,
$
g_{\boldsymbol{k},\boldsymbol{k}',\boldsymbol{q}}\,
=
\tilde{V}_{\boldsymbol{k}'-\boldsymbol{k}}
F_{\boldsymbol{k}+\frac{\boldsymbol{q}}{2},\boldsymbol{k}'+\frac{\boldsymbol{q}}{2}}
F_{-\boldsymbol{k}+\frac{\boldsymbol{q}}{2},-\boldsymbol{k}'+\frac{\boldsymbol{q}}{2}}
$.
\\
\\
Next, we introduce the mean-field,
\begin{equation}
b_{\boldsymbol{k'},\boldsymbol{Q}} \equiv \bigl\langle
c_{-\boldsymbol{k'} + \frac{\boldsymbol{q}}{2}}\, c_{\boldsymbol{k'} + \frac{\boldsymbol{q}}{2}}
\bigr\rangle,
\end{equation}
which produces the following mean-field Hamiltonian,
\begin{equation}
H_{\text{MF}} =
\sum_{\boldsymbol{k}}    
\varepsilon_{\boldsymbol{k}}
\,
c^\dagger_{\boldsymbol{k}}c_{\boldsymbol{k}}
-\frac{1}{2}
\sum_{\boldsymbol{k}}
\left\{
\Delta_{\boldsymbol{k},\boldsymbol{q}}^{*}
c_{-\boldsymbol{k} + \frac{\boldsymbol{q}}{2}}
c_{\boldsymbol{k} + \frac{\boldsymbol{q}}{2}}
+
\Delta_{\boldsymbol{k},\boldsymbol{q}}
c_{\boldsymbol{k} + \frac{\boldsymbol{q}}{2}}^\dagger
c_{-\boldsymbol{k} + \frac{\boldsymbol{q}}{2}}^\dagger
\right\}
+
K_{0,\boldsymbol{q}}
\end{equation}
with
\begin{equation}
K_{0,\boldsymbol{q}}
=
\frac{1}{2}
\sum_{\boldsymbol{k}}
\Delta_{\boldsymbol{k},\boldsymbol{q}}b_{\boldsymbol{k},\boldsymbol{q}}^{*}.
\end{equation}
In the expression for the mean-field Hamiltonian, we have defined the superconducting order parameter as,  
\begin{equation}
\Delta_{\boldsymbol{k},\boldsymbol{q}}
=
-\frac{1}{\Omega}
\sum_{\boldsymbol{k'}}
g_{\boldsymbol{k},\boldsymbol{k'},\boldsymbol{q}}
b_{\boldsymbol{k'},\boldsymbol{q}}.
\end{equation}
We note that $g_{\boldsymbol{k},\boldsymbol{k}',\boldsymbol{q}}=g_{-\boldsymbol{k},-\boldsymbol{k}',\boldsymbol{q}}$ and $b_{\boldsymbol{k},\boldsymbol{Q}}=-b_{-\boldsymbol{k},\boldsymbol{Q}}$ from the fermionic anticommutation relations. Hence, the order parameter needs to satisfy, 
\begin{equation}
\Delta_{\boldsymbol{k},\boldsymbol{q}} 
=
-
\Delta_{-\boldsymbol{k},\boldsymbol{q}}.
\end{equation}

\section{Bogoliubov-de Gennes Hamiltonian}
Here, we provide details on the derivation of the 
Bogoliubov-de Gennes Hamiltonian for rhombohedral tetralayer graphene and its diagonalization procedure.

\subsection{Hamiltonian formulation}
As a starting point, we write down the full system Hamiltonian as,
\begin{equation}
H_{\text{MF},\boldsymbol{q}} = H_{0} + H_{\Delta},
\end{equation}
where $H_{0}$ represents the normal-state Hamiltonian and $H_{\Delta}$ is the pairing Hamiltonian.
\\
\\
First, for the normal-state Hamiltonian, we write,
\begin{equation}
\begin{split}
H_{0} 
&= 
\sum_{\boldsymbol{k}} \varepsilon_{\boldsymbol{k}} c_{\boldsymbol{k}}^\dagger c_{\boldsymbol{k}} 
\\
&= 
\sum_{\boldsymbol{k}\in\text{BZ}_{+}} \varepsilon_{\boldsymbol{k}+\frac{\boldsymbol{q}}{2}} c_{\boldsymbol{k}+\frac{\boldsymbol{q}}{2}}^\dagger c_{\boldsymbol{k}+\frac{\boldsymbol{q}}{2}}
+ 
\sum_{\boldsymbol{k}\in\text{BZ}_{+}} \varepsilon_{-\boldsymbol{k}+\frac{\boldsymbol{q}}{2}} 
c_{-\boldsymbol{k}+\frac{\boldsymbol{q}}{2}} 
c_{-\boldsymbol{k}+\frac{\boldsymbol{q}}{2}}^\dagger
+ 
K_{1,\boldsymbol{q}},
\end{split}
\end{equation}
where the dispersion $\varepsilon_{\boldsymbol{k}}$ corresponds to the lowest conduction band. In this expression, the Brillouin zone is partitioned as
$
\text{BZ}_{+} \equiv \left\{\boldsymbol{k}=(k_{x},k_{y})\in\text{BZ}\,\middle|\, k_{x}>0\right\}
$
and
$
\text{BZ}_{-} \equiv \left\{\boldsymbol{k}=(k_{x},k_{y})\in\text{BZ}\,\middle|\, k_{x}<0\right\}
$. 
Moreover, the energy offset is defined as
\begin{equation}
K_{1,\boldsymbol{q}} \equiv \sum_{\boldsymbol{k}\in\text{BZ}_{+}} \varepsilon_{-\boldsymbol{k}+\frac{\boldsymbol{q}}{2}}.
\end{equation}
Second, the pairing Hamiltonian is written as
\begin{equation}
\begin{split}
H_{\Delta} &= K_{0,\boldsymbol{q}} - \frac{1}{2}\sum_{\boldsymbol{k}\in\text{BZ}_{+}}
\left\{
\Delta_{\boldsymbol{k},\boldsymbol{q}}\, c_{\boldsymbol{k}+\frac{\boldsymbol{q}}{2}}^\dagger c_{-\boldsymbol{k}+\frac{\boldsymbol{q}}{2}}^\dagger
+ \Delta_{-\boldsymbol{k},\boldsymbol{q}}\, c_{-\boldsymbol{k}+\frac{\boldsymbol{q}}{2}}^\dagger c_{\boldsymbol{k}+\frac{\boldsymbol{q}}{2}}^\dagger
+ \text{h.c.}
\right\} \\
&= K_{0,\boldsymbol{q}} - \sum_{\boldsymbol{k}\in\text{BZ}_{+}}
\left\{
\Delta_{\boldsymbol{k},\boldsymbol{q}}\, c_{\boldsymbol{k}+\frac{\boldsymbol{q}}{2}}^\dagger c_{-\boldsymbol{k}+\frac{\boldsymbol{q}}{2}}^\dagger
+ \text{h.c.}
\right\},
\end{split}
\end{equation}
Hence, the full Hamiltonian can be written as 
\begin{equation}
H = K_{0,\boldsymbol{q}} + K_{1,\boldsymbol{q}} + \sum_{\boldsymbol{k}\in\text{BZ}_{+}}
\Psi_{\boldsymbol{k},\boldsymbol{q}}^\dagger 
\mathcal{H}_{\text{BdG},\boldsymbol{k},\boldsymbol{q}}  \Psi_{\boldsymbol{k},\boldsymbol{q}},
\quad\text{with}\quad
\mathcal{H}_{\text{BdG},\boldsymbol{k},\boldsymbol{q}} \equiv
\begin{pmatrix}
\varepsilon_{\boldsymbol{k}+\frac{\boldsymbol{q}}{2}} & -\Delta_{\boldsymbol{k},\boldsymbol{q}} \\
-\Delta_{\boldsymbol{k},\boldsymbol{q}}^* & -\varepsilon_{-\boldsymbol{k}+\frac{\boldsymbol{q}}{2}}
\end{pmatrix}
\end{equation}
and the Nambu spinor, 
$
\Psi_{\boldsymbol{k},\boldsymbol{q}} 
\equiv 
( 
c_{\boldsymbol{k}+\frac{\boldsymbol{q}}{2}},
c_{-\boldsymbol{k}+\frac{\boldsymbol{q}}{2}}^\dagger 
)^T.
$

\subsection{Hamiltonian Diagonalization}
To diagonalize 
$\mathcal{H}_{\text{BdG},\boldsymbol{k},\boldsymbol{q}}$, 
we first introduce the dispersions
\begin{equation}
\xi_{\boldsymbol{k},\boldsymbol{q},\pm} 
=
\frac{
\varepsilon_{\boldsymbol{k}+\frac{\boldsymbol{q}}{2}} 
\pm 
\varepsilon_{-\boldsymbol{k}+\frac{\boldsymbol{q}}{2}}
}{2},    
\end{equation}
with
$
\xi_{\boldsymbol{k},\boldsymbol{q},+} 
=
\xi_{-\boldsymbol{k},\boldsymbol{q},+}
$
and
$
\xi_{\boldsymbol{k},\boldsymbol{q},-} 
=
-\xi_{-\boldsymbol{k},\boldsymbol{q},-}
$.
We can then express the Bogoliubov-de Gennes Hamiltonian as,
\begin{equation}
\mathcal{H}_{\text{BdG},\boldsymbol{k},\boldsymbol{q}} 
=
\xi_{\boldsymbol{k},\boldsymbol{q},-}\, \boldsymbol{I}_{2\times 2}
+
\begin{pmatrix}
\xi_{\boldsymbol{k},\boldsymbol{q},+} & 
-\Delta_{\boldsymbol{k},\boldsymbol{q}} \\
-\Delta_{\boldsymbol{k},\boldsymbol{q}}^* & 
-\xi_{\boldsymbol{k},\boldsymbol{q},+}
\end{pmatrix}.
\end{equation}
Next, we define the Bogoliubov transformation
\begin{equation}
\begin{pmatrix}
c_{\boldsymbol{k}+\frac{\boldsymbol{q}}{2}} \\
c_{-\boldsymbol{k}+\frac{\boldsymbol{q}}{2}}^\dagger
\end{pmatrix}
=
W_{\boldsymbol{k},\boldsymbol{q}}
\begin{pmatrix}
\gamma_{\boldsymbol{k}} \\
\gamma_{-\boldsymbol{k}}^\dagger
\end{pmatrix}
\quad\text{with}\quad
W_{\boldsymbol{k},\boldsymbol{q}} 
=
\begin{pmatrix}
u_{\boldsymbol{k},\boldsymbol{q}} & e^{i\phi_{\boldsymbol{k},\boldsymbol{q}}} v_{\boldsymbol{k},\boldsymbol{q}} \\[2mm]
- e^{-i\phi_{\boldsymbol{k},\boldsymbol{q}}} v_{\boldsymbol{k},\boldsymbol{q}} & u_{\boldsymbol{k},\boldsymbol{q}}
\end{pmatrix}.
\end{equation}
The phase $\phi_{\boldsymbol{k},\boldsymbol{q}}$ is defined by
$
\Delta_{\boldsymbol{k},\boldsymbol{q}} 
=
\left|\Delta_{\boldsymbol{k},\boldsymbol{q}}\right| 
e^{i\phi_{\boldsymbol{k},\boldsymbol{q}}}
$. Moreover, the coherence factors are given by
\begin{equation}
\begin{split}
u_{\boldsymbol{k},\boldsymbol{q}}^2 
&=
\frac{1}{2}
\left( 
1 + \frac{\xi_{\boldsymbol{k},\boldsymbol{q},+}}{\sqrt{\xi_{\boldsymbol{k},\boldsymbol{q},+}^2 
+ |\Delta_{\boldsymbol{k},\boldsymbol{q}}|^2}} 
\right), 
\qquad
v_{\boldsymbol{k},\boldsymbol{q}}^2 
=
\frac{1}{2}
\left( 
1 - \frac{\xi_{\boldsymbol{k},\boldsymbol{q},+}}{\sqrt{\xi_{\boldsymbol{k},\boldsymbol{q},+}^2 
+ |\Delta_{\boldsymbol{k},\boldsymbol{q}}|^2}} 
\right),
\\
u_{\boldsymbol{k},\boldsymbol{q}}\,v_{\boldsymbol{k},\boldsymbol{q}} 
&=
\frac{|\Delta_{\boldsymbol{k},\boldsymbol{q}}|}{2\sqrt{\xi_{\boldsymbol{k},\boldsymbol{q},+}^2 
+ |\Delta_{\boldsymbol{k},\boldsymbol{q}}|^2}}.
\end{split}
\end{equation}
With this transformation, the Bogoliubov-de Gennes Hamiltonian becomes diagonal,
\begin{equation}
W_{\boldsymbol{k},\boldsymbol{q}}^\dagger
\mathcal{H}_{\text{BdG},\boldsymbol{k},\boldsymbol{q}}
W_{\boldsymbol{k},\boldsymbol{q}} 
=
\begin{pmatrix}
E_{\boldsymbol{k},\boldsymbol{q},+} & 0 \\
0 & -E_{\boldsymbol{k},\boldsymbol{q},-}
\end{pmatrix},
\end{equation}
where the energy dispersions are given by
\begin{equation}
E_{\boldsymbol{k},\boldsymbol{q},\pm} 
=
\pm \xi_{\boldsymbol{k},\boldsymbol{q},-} 
+
\sqrt{\xi_{\boldsymbol{k},\boldsymbol{q},+}^2 + |\Delta_{\boldsymbol{k},\boldsymbol{q}}|^2}.
\end{equation}
Moreover, we define,
\begin{equation}
E_{\boldsymbol{k},\boldsymbol{q}} 
\equiv \sqrt{\xi_{\boldsymbol{k},\boldsymbol{q},+}^2 + |\Delta_{\boldsymbol{k},\boldsymbol{q}}|^2}.
\end{equation}
The full system Hamiltonian then takes on the diagonalized form
\begin{equation}
\begin{split}
H 
&=
K_{0,\boldsymbol{q}} + K_{1,\boldsymbol{q}} 
+
\sum_{\boldsymbol{k}\in\text{BZ}_{+}}
[ 
E_{\boldsymbol{k},\boldsymbol{q},+}
\gamma_{\boldsymbol{k}}^\dagger\gamma_{\boldsymbol{k}}
- 
E_{\boldsymbol{k},\boldsymbol{q},-}
\gamma_{-\boldsymbol{k}}\gamma_{-\boldsymbol{k}}^\dagger 
] 
\\
&=
K_{0,\boldsymbol{q}} 
+
K_{1,\boldsymbol{q}} 
+
K_{2,\boldsymbol{q}} 
+ 
\sum_{\boldsymbol{k}\in\text{BZ}_{+}}
[ 
E_{\boldsymbol{k},\boldsymbol{q},+}
\gamma_{\boldsymbol{k}}^\dagger\gamma_{\boldsymbol{k}}
+ E_{\boldsymbol{k},\boldsymbol{q},-}
\gamma_{-\boldsymbol{k}}^\dagger\gamma_{-\boldsymbol{k}} 
] 
\\
&=
K_{\boldsymbol{q}} 
+
\sum_{\boldsymbol{k}\in\text{BZ}_{+}}
E_{\boldsymbol{k},\boldsymbol{q},+}
\gamma_{\boldsymbol{k}}^\dagger\gamma_{\boldsymbol{k}}
+ 
\sum_{\boldsymbol{k}\in\text{BZ}_{-}}
E_{-\boldsymbol{k},\boldsymbol{q},-}
\gamma_{\boldsymbol{k}}^\dagger\gamma_{\boldsymbol{k}} 
\\
&=
K_{\boldsymbol{q}} 
+
\sum_{\boldsymbol{k}\in\text{BZ}_{+}}
E_{\boldsymbol{k},\boldsymbol{q},+}
\gamma_{\boldsymbol{k}}^\dagger\gamma_{\boldsymbol{k}}
+ 
\sum_{\boldsymbol{k}\in\text{BZ}_{-}}
E_{\boldsymbol{k},\boldsymbol{q},+}
\gamma_{\boldsymbol{k}}^\dagger\gamma_{\boldsymbol{k}}
\\
&=
K_{\boldsymbol{q}} 
+
\sum_{\boldsymbol{k}\in\text{BZ}}
E_{\boldsymbol{k},\boldsymbol{q},+}
\gamma_{\boldsymbol{k}}^\dagger\gamma_{\boldsymbol{k}},
\end{split}
\end{equation}
where we have used that
$
E_{\boldsymbol{k},\boldsymbol{q},+} 
=
E_{-\boldsymbol{k},\boldsymbol{q},-},
$
and defined the energy offset
\begin{equation}
K_{2,\boldsymbol{q}} 
=
-\sum_{\boldsymbol{k}\in\text{BZ}_{+}} E_{\boldsymbol{k},\boldsymbol{q},-}.
\end{equation}
The total energy offset is given by
\begin{equation}
\begin{split}
K_{\boldsymbol{q}} 
&=
K_{0,\boldsymbol{q}} 
+
K_{1,\boldsymbol{q}} 
+
K_{2,\boldsymbol{q}} 
\\
&=
\sum_{\boldsymbol{k}} 
\frac{1}{2}
\Delta_{\boldsymbol{k},\boldsymbol{q}}
b_{\boldsymbol{k},\boldsymbol{q}}^*
+ 
\sum_{\boldsymbol{k}\in\text{BZ}_{+}} 
\left\{ 
\varepsilon_{-\boldsymbol{k}+\frac{\boldsymbol{q}}{2}} 
-
E_{\boldsymbol{k},\boldsymbol{q},-} \right\}
\\
&=
\sum_{\boldsymbol{k}} \frac{1}{2}\Delta_{\boldsymbol{k},\boldsymbol{q}}\,b_{\boldsymbol{k},\boldsymbol{q}}^*
+ \frac{1}{2}\sum_{\boldsymbol{k}\in\text{BZ}} 
\Bigl\{ \xi_{\boldsymbol{k},\boldsymbol{q},+} - E_{\boldsymbol{k},\boldsymbol{q}} \Bigr\}.
\end{split}
\end{equation}
This completes the formulation and diagonalization of the Bogoliubov-de Gennes Hamiltonian for rhombohedral tetralayer graphene.

\section{Self-consistency Equation for the Superconducting Gap Function}

Here, we provide details on the derivation of the self-consistency equation that determines the superconducting gap function. 
\\
\\
For this purpose, we need to evaluate the pair correlation function
$
b_{\boldsymbol{k},\boldsymbol{q}} 
=
\langle 
c_{-\boldsymbol{k}+\frac{\boldsymbol{q}}{2}} c_{\boldsymbol{k}+\frac{\boldsymbol{q}}{2}}
\rangle.
$
\\
\\
\textit{As a first step}, assume that $\boldsymbol{k}\in\text{BZ}_{+}$. Then we can write
\begin{equation}
\begin{split}
c_{-\boldsymbol{k}+\frac{\boldsymbol{q}}{2}} c_{\boldsymbol{k}+\frac{\boldsymbol{q}}{2}}
&=
\left\{
 - e^{i\phi_{\boldsymbol{k},\boldsymbol{q}}} v_{\boldsymbol{k},\boldsymbol{q}} \gamma_{\boldsymbol{k}}^\dagger
+ u_{\boldsymbol{k},\boldsymbol{q}} 
\gamma_{-\boldsymbol{k}}
\right\} 
\left\{
 u_{\boldsymbol{k},\boldsymbol{q}} \gamma_{\boldsymbol{k}}
+ e^{i\phi_{\boldsymbol{k},\boldsymbol{q}}} v_{\boldsymbol{k},\boldsymbol{q}} \gamma_{-\boldsymbol{k}}^\dagger
\right\} \\
&=
- e^{i\phi_{\boldsymbol{k},\boldsymbol{q}}} v_{\boldsymbol{k},\boldsymbol{q}} u_{\boldsymbol{k},\boldsymbol{q}} \gamma_{\boldsymbol{k}}^\dagger \gamma_{\boldsymbol{k}} 
- e^{2i\phi_{\boldsymbol{k},\boldsymbol{q}}} v_{\boldsymbol{k},\boldsymbol{q}}^2 \gamma_{\boldsymbol{k}}^\dagger \gamma_{-\boldsymbol{k}}^\dagger 
+ u_{\boldsymbol{k},\boldsymbol{q}}^2 \gamma_{-\boldsymbol{k}} \gamma_{\boldsymbol{k}}  + e^{i\phi_{\boldsymbol{k},\boldsymbol{q}}} u_{\boldsymbol{k},\boldsymbol{q}} v_{\boldsymbol{k},\boldsymbol{q}} \gamma_{-\boldsymbol{k}} \gamma_{-\boldsymbol{k}}^\dagger.
\end{split}
\end{equation}
Next, we evaluate the following expectation values,
\begin{equation}
\begin{split}
\langle \gamma_{\boldsymbol{k}}^\dagger \gamma_{\boldsymbol{k}} \rangle 
&= n_F\left( E_{\boldsymbol{k},\boldsymbol{q},+} \right), \\
\langle \gamma_{-\boldsymbol{k}} \gamma_{-\boldsymbol{k}}^\dagger \rangle 
&= 1 - \langle \gamma_{-\boldsymbol{k}}^\dagger \gamma_{-\boldsymbol{k}} \rangle 
= 1 - n_F\left( E_{\boldsymbol{k},\boldsymbol{q},-} \right)
= 1 - n_F\left( E_{-\boldsymbol{k},\boldsymbol{q},+} \right), \\
\langle \gamma_{\boldsymbol{k}}^\dagger \gamma_{-\boldsymbol{k}}^\dagger \rangle 
&= 0, \\
\langle \gamma_{-\boldsymbol{k}} \gamma_{\boldsymbol{k}} \rangle 
&= 0.
\end{split}
\end{equation}
Here,
$
n_F(E) \equiv \frac{1}{e^{\beta E}+1}
$
is the Fermi-Dirac distribution function.
\\
\\
Using these results, the pair correlation function becomes
\begin{equation}
\begin{split}
b_{\boldsymbol{k},\boldsymbol{q}} 
&=
e^{i\phi_{\boldsymbol{k},\boldsymbol{q}}} u_{\boldsymbol{k},\boldsymbol{q}} v_{\boldsymbol{k},\boldsymbol{q}} 
\{ 1 - n_F\left( E_{\boldsymbol{k},\boldsymbol{q},+} \right)
- n_F\left( E_{-\boldsymbol{k},\boldsymbol{q},+} \right\} \\
&=
\frac{\Delta_{\boldsymbol{k},\boldsymbol{q}}}{2 E_{\boldsymbol{k},\boldsymbol{q}}} 
\{ 1 - n_F\left( E_{\boldsymbol{k},\boldsymbol{q},+} \right)
- n_F\left( E_{-\boldsymbol{k},\boldsymbol{q},+} \right\}.
\end{split}
\end{equation}
This result holds for $\boldsymbol{k}\in\text{BZ}_{+}$.
\\
\\
\textit{As a second step}, assume that $\boldsymbol{k}\in\text{BZ}_{-}$. Then we can write 
$
\boldsymbol{k} = -\boldsymbol{k}'
$
for some $\boldsymbol{k}'\in\text{BZ}_{+}$. In this case,
\begin{equation}
\begin{split}
b_{\boldsymbol{k},\boldsymbol{q}} 
&= \left\langle c_{-\boldsymbol{k}+\frac{\boldsymbol{q}}{2}} c_{\boldsymbol{k}+\frac{\boldsymbol{q}}{2}} \right\rangle 
= \left\langle c_{\boldsymbol{k}'+\frac{\boldsymbol{q}}{2}} c_{-\boldsymbol{k}'+\frac{\boldsymbol{q}}{2}} \right\rangle 
= - \left\langle c_{-\boldsymbol{k}'+\frac{\boldsymbol{q}}{2}} c_{\boldsymbol{k}'+\frac{\boldsymbol{q}}{2}} \right\rangle 
= -b_{\boldsymbol{k}',\boldsymbol{q}} \\
&= -\frac{\Delta_{\boldsymbol{k}',\boldsymbol{q}}}{2 E_{\boldsymbol{k}',\boldsymbol{q}}} 
\left\{ 1 - n_F\left( E_{\boldsymbol{k}',\boldsymbol{q},+} \right)
- n_F\left( E_{-\boldsymbol{k}',\boldsymbol{q},+} \right) \right\} \\
&= \frac{\Delta_{\boldsymbol{k},\boldsymbol{q}}}{2 E_{\boldsymbol{k},\boldsymbol{q}}} 
\left\{ 1 - n_F\left( E_{\boldsymbol{k},\boldsymbol{q},+} \right)
- n_F\left( E_{-\boldsymbol{k},\boldsymbol{q},+} \right) \right\}.
\end{split}
\end{equation}
Note that in the first four equalities we are only using the definition of the pair correlation function and the fermionic anti-commutation relations. Only in the fifth equality, we use that $\boldsymbol{k}'\in\text{BZ}_+$. In the sixth equality, we use the that the superconducting order parameter is an odd function for all 
$\boldsymbol{k}\in\text{BZ}$.
\\
\\
Thus, the following expression for the pair correlation function holds for all $\boldsymbol{k}\in\text{BZ}$,
\begin{equation}
b_{\boldsymbol{k},\boldsymbol{q}} 
=
\frac{\Delta_{\boldsymbol{k},\boldsymbol{q}}}{2 E_{\boldsymbol{k},\boldsymbol{q}}} 
\left\{ 1 - n_F\left( E_{\boldsymbol{k},\boldsymbol{q},+} \right)
- n_F\left( E_{-\boldsymbol{k},\boldsymbol{q},+} \right) \right\}.
\end{equation}
\\
\\
As a result, the self-consistency gap equation is given by,
\begin{equation}
\begin{split}
\Delta_{\boldsymbol{k},\boldsymbol{q}} 
&=
- \sum_{\boldsymbol{k}'} 
g_{\boldsymbol{k},\boldsymbol{k'},\boldsymbol{q}} b_{\boldsymbol{k}',\boldsymbol{q}} \\
&=
- \sum_{\boldsymbol{k}'} 
g_{\boldsymbol{k},\boldsymbol{k'},\boldsymbol{q}} 
 \frac{\Delta_{\boldsymbol{k}',\boldsymbol{q}}}{2 E_{\boldsymbol{k}',\boldsymbol{q}}} \left\{ 1 - n_F\left( E_{\boldsymbol{k}',\boldsymbol{q},+} \right)
- n_F\left( E_{-\boldsymbol{k}',\boldsymbol{q},+} \right) \right\} \\
&=
- \sum_{\boldsymbol{k}'} 
g_{\boldsymbol{k},\boldsymbol{k'},\boldsymbol{q}} 
\frac{\Delta_{\boldsymbol{k}',\boldsymbol{q}}}{2 E_{\boldsymbol{k}',\boldsymbol{q}}} \, \frac{1}{2} \left[\tanh\left( \frac{E_{\boldsymbol{k}',\boldsymbol{q},+}}{2 k_{B}T} \right)
+\tanh\left( \frac{E_{-\boldsymbol{k}',\boldsymbol{q},+}}{2 k_{B}T} \right)\right].
\end{split}
\end{equation}

\section{Total Electron Density}
Here, we provide details on the derivation of the total electron density in the superconducting state of the rhombohedral tetralayer graphene system. 
\\
\\
In the our mean-field approach to superconductivity, the occupancy of a particular momentum state depends on both 
the superconducting gap function, $\Delta_{\boldsymbol{k},\boldsymbol{Q}}$, and
the chemical potential, $\mu$. As a result, the total electron density,
\begin{equation}
n = \frac{1}{\Omega} \sum_{\boldsymbol{k}} \langle c_{\boldsymbol{k}}^\dagger c_{\boldsymbol{k}} \rangle.
\end{equation}
\\
\\
To compute the total electron density, let us write,
\begin{equation}
\begin{split}
n 
= \frac{1}{\Omega}\sum_{\boldsymbol{k}} \langle c_{\boldsymbol{k}}^\dagger c_{\boldsymbol{k}} \rangle
= 
\frac{1}{\Omega}\sum_{\boldsymbol{k}} \langle c_{\boldsymbol{k}+\frac{\boldsymbol{q}}{2}}^\dagger c_{\boldsymbol{k}+\frac{\boldsymbol{q}}{2}} \rangle 
=
\frac{1}{\Omega}\sum_{\boldsymbol{k}\in\text{BZ}_{+}} \langle c_{\boldsymbol{k}+\frac{\boldsymbol{q}}{2}}^\dagger c_{\boldsymbol{k}+\frac{\boldsymbol{q}}{2}} \rangle 
+ \frac{1}{\Omega}\sum_{\boldsymbol{k}\in\text{BZ}_{+}} \langle c_{-\boldsymbol{k}+\frac{\boldsymbol{q}}{2}}^\dagger c_{-\boldsymbol{k}+\frac{\boldsymbol{q}}{2}} \rangle.
\end{split}
\end{equation}
For a given $\boldsymbol{k}\in\text{BZ}_{+}$, these expectation values are evaluated as,
\begin{equation}
\begin{split}
\langle c_{\boldsymbol{k}+\frac{\boldsymbol{q}}{2}}^\dagger c_{\boldsymbol{k}+\frac{\boldsymbol{q}}{2}} \rangle 
&=
u_{\boldsymbol{k},\boldsymbol{q}}^2 n_F\left( E_{\boldsymbol{k},\boldsymbol{q},+} \right)
- v_{\boldsymbol{k},\boldsymbol{q}}^2 n_F\left( E_{-\boldsymbol{k},\boldsymbol{q},+} \right)
+ v_{\boldsymbol{k},\boldsymbol{q}}^2, \\[2mm]
\langle c_{-\boldsymbol{k}+\frac{\boldsymbol{q}}{2}}^\dagger c_{-\boldsymbol{k}+\frac{\boldsymbol{q}}{2}} \rangle 
&=
u_{\boldsymbol{k},\boldsymbol{q}}^2 n_F\left( E_{-\boldsymbol{k},\boldsymbol{q},+} \right)
- v_{\boldsymbol{k},\boldsymbol{q}}^2 n_F\left( E_{\boldsymbol{k},\boldsymbol{q},+} \right)
+ v_{\boldsymbol{k},\boldsymbol{q}}^2.
\end{split}
\end{equation}
Using that
$u_{\boldsymbol{k},\boldsymbol{q}}^2 = u_{-\boldsymbol{k},\boldsymbol{q}}^2$ and $v_{\boldsymbol{k},\boldsymbol{q}}^2 = v_{-\boldsymbol{k},\boldsymbol{q}}^2$,
we obtain the total electron density as
\begin{equation}
n = \frac{1}{\Omega}\sum_{\boldsymbol{k}\in\text{BZ}}
\left\{
u_{\boldsymbol{k},\boldsymbol{q}}^2 n_F\left( E_{\boldsymbol{k},\boldsymbol{q},+} \right)
- v_{\boldsymbol{k},\boldsymbol{q}}^2 n_F\left( E_{-\boldsymbol{k},\boldsymbol{q},+} \right)
+ v_{\boldsymbol{k},\boldsymbol{q}}^2
\right\}.
\end{equation}

\section{Free Energy}
In this section of the Supplemental Material, we provide details on the derivation of the free energy of the rhombohedral tetralayer graphene system.
\\
\\
First, we evaluate the partition function,
\begin{equation}
\begin{split}
Z &= \mathrm{Tr}\left[e^{-\beta H}\right] \\
  &= e^{-\beta K_{\boldsymbol{q}}} \prod_{\boldsymbol{k}\in\mathrm{BZ}_{+}} 
  \left\{ 1 + e^{-\beta E_{\boldsymbol{k},\boldsymbol{q},+}} \right\}
  \left\{ 1 + e^{-\beta E_{-\boldsymbol{k},\boldsymbol{q},+}} \right\} \\
  &= e^{-\beta K_{\boldsymbol{q}}} \prod_{\boldsymbol{k}\in\mathrm{BZ}} 
  \left\{ 1 + e^{-\beta E_{\boldsymbol{k},\boldsymbol{q},+}} \right\}.
\end{split}
\end{equation}
Then the free energy is obtained as,
\begin{equation}
\begin{split}
\mathcal{F} &= -\frac{1}{\beta}\ln Z \\
&= K_{\boldsymbol{q}} - \frac{1}{\beta} \sum_{\boldsymbol{k}\in\mathrm{BZ}_{+}} \ln\left[ 1 + e^{-\beta E_{\boldsymbol{k},\boldsymbol{q},+}} \right] 
- \frac{1}{\beta} \sum_{\boldsymbol{k}\in\mathrm{BZ}_{+}} \ln\left[ 1 + e^{-\beta E_{-\boldsymbol{k},\boldsymbol{q},+}} \right] \\
&= K_{\boldsymbol{q}} - \frac{1}{\beta} \sum_{\boldsymbol{k}\in\mathrm{BZ}} \ln\left[ 1 + e^{-\beta E_{\boldsymbol{k},\boldsymbol{q},+}} \right].
\end{split}
\end{equation}
For convenience, let us write down again the energy offset,
\begin{equation}
K_{\boldsymbol{q}} 
=
\sum_{\boldsymbol{k}} \frac{1}{2}\Delta_{\boldsymbol{k},\boldsymbol{q}}\,b_{\boldsymbol{k},\boldsymbol{q}}^*
+ \frac{1}{2}\sum_{\boldsymbol{k}} 
\{ \xi_{\boldsymbol{k},\boldsymbol{q},+} - E_{\boldsymbol{k},\boldsymbol{q}} \}
\end{equation}
where
\begin{equation}
b_{\boldsymbol{k},\boldsymbol{q}} 
=
\frac{\Delta_{\boldsymbol{k},\boldsymbol{q}}}{2 E_{\boldsymbol{k},\boldsymbol{q}}} 
\left\{ 1 - n_F\left( E_{\boldsymbol{k},\boldsymbol{q},+} \right)
- n_F\left( E_{-\boldsymbol{k},\boldsymbol{q},+} \right) \right\}.
\end{equation}

\end{widetext}


\begin{thebibliography}{66}
\expandafter\ifx\csname natexlab\endcsname\relax\def\natexlab#1{#1}\fi
\expandafter\ifx\csname bibnamefont\endcsname\relax
  \def\bibnamefont#1{#1}\fi
\expandafter\ifx\csname bibfnamefont\endcsname\relax
  \def\bibfnamefont#1{#1}\fi
\expandafter\ifx\csname citenamefont\endcsname\relax
  \def\citenamefont#1{#1}\fi
\expandafter\ifx\csname url\endcsname\relax
  \def\url#1{\texttt{#1}}\fi
\expandafter\ifx\csname urlprefix\endcsname\relax\def\urlprefix{URL }\fi
\providecommand{\bibinfo}[2]{#2}
\providecommand{\eprint}[2][]{\url{#2}}

\bibitem[{\citenamefont{Nadeem et~al.}(2023)\citenamefont{Nadeem, Fuhrer, and Wang}}]{nadeem2023superconducting}
\bibinfo{author}{\bibfnamefont{M.}~\bibnamefont{Nadeem}}, \bibinfo{author}{\bibfnamefont{M.~S.} \bibnamefont{Fuhrer}}, \bibnamefont{and} \bibinfo{author}{\bibfnamefont{X.}~\bibnamefont{Wang}}, \href{https://www.nature.com/articles/s42254-023-00632-w}{\bibinfo{title}{The superconducting diode effect}}, \bibinfo{journal}{Nature Reviews Physics} \textbf{\bibinfo{volume}{5}}, \bibinfo{pages}{558} (\bibinfo{year}{2023}).

\bibitem[{\citenamefont{Ando et~al.}(2020)\citenamefont{Ando, Miyasaka, Li, Ishizuka, Arakawa, Shiota, Moriyama, Yanase, and Ono}}]{ando_observation_2020}
\bibinfo{author}{\bibfnamefont{F.}~\bibnamefont{Ando}}, \bibinfo{author}{\bibfnamefont{Y.}~\bibnamefont{Miyasaka}}, \bibinfo{author}{\bibfnamefont{T.}~\bibnamefont{Li}}, \bibinfo{author}{\bibfnamefont{J.}~\bibnamefont{Ishizuka}}, \bibinfo{author}{\bibfnamefont{T.}~\bibnamefont{Arakawa}}, \bibinfo{author}{\bibfnamefont{Y.}~\bibnamefont{Shiota}}, \bibinfo{author}{\bibfnamefont{T.}~\bibnamefont{Moriyama}}, \bibinfo{author}{\bibfnamefont{Y.}~\bibnamefont{Yanase}}, \bibnamefont{and} \bibinfo{author}{\bibfnamefont{T.}~\bibnamefont{Ono}}, \href{https://www.nature.com/articles/s41586-020-2590-4}{\bibinfo{title}{Observation of superconducting diode effect}}, \bibinfo{journal}{Nature} \textbf{\bibinfo{volume}{584}}, \bibinfo{pages}{373} (\bibinfo{year}{2020}).

\bibitem[{\citenamefont{He et~al.}(2022)\citenamefont{He, Tanaka, and Nagaosa}}]{he_phenomenological_2022}
\bibinfo{author}{\bibfnamefont{J.~J.} \bibnamefont{He}}, \bibinfo{author}{\bibfnamefont{Y.}~\bibnamefont{Tanaka}}, \bibnamefont{and} \bibinfo{author}{\bibfnamefont{N.}~\bibnamefont{Nagaosa}}, \href{https://iopscience.iop.org/article/10.1088/1367-2630/ac6766/meta}{\bibinfo{title}{A phenomenological theory of superconductor diodes}}, \bibinfo{journal}{New Journal of Physics} \textbf{\bibinfo{volume}{24}}, \bibinfo{pages}{053014} (\bibinfo{year}{2022}).

\bibitem[{\citenamefont{Yuan and Fu}(2022)}]{yuan_supercurrent_2022}
\bibinfo{author}{\bibfnamefont{N.~F.} \bibnamefont{Yuan}} \bibnamefont{and} \bibinfo{author}{\bibfnamefont{L.}~\bibnamefont{Fu}}, \href{https://www.pnas.org/doi/10.1073/pnas.2119548119}{\bibinfo{title}{Supercurrent diode effect and finite-momentum superconductors}}, \bibinfo{journal}{Proceedings of the National Academy of Sciences} \textbf{\bibinfo{volume}{119}}, \bibinfo{pages}{e2119548119} (\bibinfo{year}{2022}).

\bibitem[{\citenamefont{Zhang et~al.}(2022)\citenamefont{Zhang, Gu, Li, Hu, and Jiang}}]{zhang_general_2022}
\bibinfo{author}{\bibfnamefont{Y.}~\bibnamefont{Zhang}}, \bibinfo{author}{\bibfnamefont{Y.}~\bibnamefont{Gu}}, \bibinfo{author}{\bibfnamefont{P.}~\bibnamefont{Li}}, \bibinfo{author}{\bibfnamefont{J.}~\bibnamefont{Hu}}, \bibnamefont{and} \bibinfo{author}{\bibfnamefont{K.}~\bibnamefont{Jiang}}, \href{https://link.aps.org/doi/10.1103/PhysRevX.12.041013}{\bibinfo{title}{General {Theory} of {Josephson} {Diodes}}}, \bibinfo{journal}{Physical Review X} \textbf{\bibinfo{volume}{12}}, \bibinfo{pages}{041013} (\bibinfo{year}{2022}).

\bibitem[{\citenamefont{Misaki and Nagaosa}(2021)}]{misaki_theory_2021}
\bibinfo{author}{\bibfnamefont{K.}~\bibnamefont{Misaki}} \bibnamefont{and} \bibinfo{author}{\bibfnamefont{N.}~\bibnamefont{Nagaosa}}, \href{https://link.aps.org/doi/10.1103/PhysRevB.103.245302}{\bibinfo{title}{Theory of the nonreciprocal {Josephson} effect}}, \bibinfo{journal}{Physical Review B} \textbf{\bibinfo{volume}{103}}, \bibinfo{pages}{245302} (\bibinfo{year}{2021}).

\bibitem[{\citenamefont{Davydova et~al.}(2022)\citenamefont{Davydova, Prembabu, and Fu}}]{davydova_universal_2022}
\bibinfo{author}{\bibfnamefont{M.}~\bibnamefont{Davydova}}, \bibinfo{author}{\bibfnamefont{S.}~\bibnamefont{Prembabu}}, \bibnamefont{and} \bibinfo{author}{\bibfnamefont{L.}~\bibnamefont{Fu}}, \href{https://www.science.org/doi/full/10.1126/sciadv.abo0309}{\bibinfo{title}{Universal {Josephson} diode effect}}, \bibinfo{journal}{Science advances} \textbf{\bibinfo{volume}{8}}, \bibinfo{pages}{eabo0309} (\bibinfo{year}{2022}).

\bibitem[{\citenamefont{Souto et~al.}(2024)\citenamefont{Souto, Leijnse, Schrade, Valentini, Katsaros, and Danon}}]{souto2024tuning}
\bibinfo{author}{\bibfnamefont{R.~S.} \bibnamefont{Souto}}, \bibinfo{author}{\bibfnamefont{M.}~\bibnamefont{Leijnse}}, \bibinfo{author}{\bibfnamefont{C.}~\bibnamefont{Schrade}}, \bibinfo{author}{\bibfnamefont{M.}~\bibnamefont{Valentini}}, \bibinfo{author}{\bibfnamefont{G.}~\bibnamefont{Katsaros}}, \bibnamefont{and} \bibinfo{author}{\bibfnamefont{J.}~\bibnamefont{Danon}}, \href{https://journals.aps.org/prresearch/abstract/10.1103/PhysRevResearch.6.L022002}{\bibinfo{title}{Tuning the {Josephson} diode response with an ac current}}, \bibinfo{journal}{Physical Review Research} \textbf{\bibinfo{volume}{6}}, \bibinfo{pages}{L022002} (\bibinfo{year}{2024}).

\bibitem[{\citenamefont{Zazunov et~al.}(2009)\citenamefont{Zazunov, Egger, Jonckheere, and Martin}}]{zazunov2009anomalous}
\bibinfo{author}{\bibfnamefont{A.}~\bibnamefont{Zazunov}}, \bibinfo{author}{\bibfnamefont{R.}~\bibnamefont{Egger}}, \bibinfo{author}{\bibfnamefont{T.}~\bibnamefont{Jonckheere}}, \bibnamefont{and} \bibinfo{author}{\bibfnamefont{T.}~\bibnamefont{Martin}}, \href{https://journals.aps.org/prl/abstract/10.1103/PhysRevLett.103.147004}{\bibinfo{title}{Anomalous {Josephson} current through a spin-orbit coupled quantum dot}}, \bibinfo{journal}{Physical Review Letters} \textbf{\bibinfo{volume}{103}}, \bibinfo{pages}{147004} (\bibinfo{year}{2009}).

\bibitem[{\citenamefont{Brunetti et~al.}(2013)\citenamefont{Brunetti, Zazunov, Kundu, and Egger}}]{brunetti2013anomalous}
\bibinfo{author}{\bibfnamefont{A.}~\bibnamefont{Brunetti}}, \bibinfo{author}{\bibfnamefont{A.}~\bibnamefont{Zazunov}}, \bibinfo{author}{\bibfnamefont{A.}~\bibnamefont{Kundu}}, \bibnamefont{and} \bibinfo{author}{\bibfnamefont{R.}~\bibnamefont{Egger}}, \href{https://journals.aps.org/prb/abstract/10.1103/PhysRevB.88.144515}{\bibinfo{title}{Anomalous {Josephson} current, incipient time-reversal symmetry breaking, and {Majorana} bound states in interacting multilevel dots}}, \bibinfo{journal}{Physical Review B} \textbf{\bibinfo{volume}{88}}, \bibinfo{pages}{144515} (\bibinfo{year}{2013}).

\bibitem[{\citenamefont{Pal et~al.}(2022)\citenamefont{Pal, Chakraborty, Sivakumar, Davydova, Gopi, Pandeya, Krieger, Zhang, Date, Ju et~al.}}]{pal2022josephson}
\bibinfo{author}{\bibfnamefont{B.}~\bibnamefont{Pal}}, \bibinfo{author}{\bibfnamefont{A.}~\bibnamefont{Chakraborty}}, \bibinfo{author}{\bibfnamefont{P.~K.} \bibnamefont{Sivakumar}}, \bibinfo{author}{\bibfnamefont{M.}~\bibnamefont{Davydova}}, \bibinfo{author}{\bibfnamefont{A.~K.} \bibnamefont{Gopi}}, \bibinfo{author}{\bibfnamefont{A.~K.} \bibnamefont{Pandeya}}, \bibinfo{author}{\bibfnamefont{J.~A.} \bibnamefont{Krieger}}, \bibinfo{author}{\bibfnamefont{Y.}~\bibnamefont{Zhang}}, \bibinfo{author}{\bibfnamefont{M.}~\bibnamefont{Date}}, \bibinfo{author}{\bibfnamefont{S.}~\bibnamefont{Ju}}, \bibnamefont{et~al.}, \href{https://www.nature.com/articles/s41567-022-01699-5}{\bibinfo{title}{Josephson diode effect from {Cooper} pair momentum in a topological semimetal}}, \bibinfo{journal}{Nature Physics} \textbf{\bibinfo{volume}{18}}, \bibinfo{pages}{1228} (\bibinfo{year}{2022}).

\bibitem[{\citenamefont{Baumgartner et~al.}(2022)\citenamefont{Baumgartner, Fuchs, Costa, Reinhardt, Gronin, Gardner, Lindemann, Manfra, Faria~Junior, Kochan et~al.}}]{baumgartner_supercurrent_2022}
\bibinfo{author}{\bibfnamefont{C.}~\bibnamefont{Baumgartner}}, \bibinfo{author}{\bibfnamefont{L.}~\bibnamefont{Fuchs}}, \bibinfo{author}{\bibfnamefont{A.}~\bibnamefont{Costa}}, \bibinfo{author}{\bibfnamefont{S.}~\bibnamefont{Reinhardt}}, \bibinfo{author}{\bibfnamefont{S.}~\bibnamefont{Gronin}}, \bibinfo{author}{\bibfnamefont{G.~C.} \bibnamefont{Gardner}}, \bibinfo{author}{\bibfnamefont{T.}~\bibnamefont{Lindemann}}, \bibinfo{author}{\bibfnamefont{M.~J.} \bibnamefont{Manfra}}, \bibinfo{author}{\bibfnamefont{P.~E.} \bibnamefont{Faria~Junior}}, \bibinfo{author}{\bibfnamefont{D.}~\bibnamefont{Kochan}}, \bibnamefont{et~al.}, \href{https://www.nature.com/articles/s41565-021-01009-9}{\bibinfo{title}{Supercurrent rectification and magnetochiral effects in symmetric {Josephson} junctions}}, \bibinfo{journal}{Nature Nanotechnology} \textbf{\bibinfo{volume}{17}}, \bibinfo{pages}{39} (\bibinfo{year}{2022}).

\bibitem[{\citenamefont{Legg et~al.}(2022)\citenamefont{Legg, Loss, and Klinovaja}}]{legg2022superconducting}
\bibinfo{author}{\bibfnamefont{H.~F.} \bibnamefont{Legg}}, \bibinfo{author}{\bibfnamefont{D.}~\bibnamefont{Loss}}, \bibnamefont{and} \bibinfo{author}{\bibfnamefont{J.}~\bibnamefont{Klinovaja}}, \href{https://journals.aps.org/prb/abstract/10.1103/PhysRevB.106.104501}{\bibinfo{title}{Superconducting diode effect due to magnetochiral anisotropy in topological insulators and {Rashba} nanowires}}, \bibinfo{journal}{Physical Review B} \textbf{\bibinfo{volume}{106}}, \bibinfo{pages}{104501} (\bibinfo{year}{2022}).

\bibitem[{\citenamefont{Lotfizadeh et~al.}(2024)\citenamefont{Lotfizadeh, Schiela, Pekerten, Yu, Elfeky, Strickland, Matos-Abiague, and Shabani}}]{lotfizadeh2023superconducting}
\bibinfo{author}{\bibfnamefont{N.}~\bibnamefont{Lotfizadeh}}, \bibinfo{author}{\bibfnamefont{W.~F.} \bibnamefont{Schiela}}, \bibinfo{author}{\bibfnamefont{B.}~\bibnamefont{Pekerten}}, \bibinfo{author}{\bibfnamefont{P.}~\bibnamefont{Yu}}, \bibinfo{author}{\bibfnamefont{B.~H.} \bibnamefont{Elfeky}}, \bibinfo{author}{\bibfnamefont{W.~M.} \bibnamefont{Strickland}}, \bibinfo{author}{\bibfnamefont{A.}~\bibnamefont{Matos-Abiague}}, \bibnamefont{and} \bibinfo{author}{\bibfnamefont{J.}~\bibnamefont{Shabani}}, \href{https://www.nature.com/articles/s42005-024-01618-5}{\bibinfo{title}{Superconducting diode effect sign change in epitaxial {Al}-{InAs} {Josephson} junctions}}, \bibinfo{journal}{Communications Physics} \textbf{\bibinfo{volume}{7}}, \bibinfo{pages}{120} (\bibinfo{year}{2024}).

\bibitem[{\citenamefont{Costa et~al.}(2023)\citenamefont{Costa, Fabian, and Kochan}}]{costa2023microscopic}
\bibinfo{author}{\bibfnamefont{A.}~\bibnamefont{Costa}}, \bibinfo{author}{\bibfnamefont{J.}~\bibnamefont{Fabian}}, \bibnamefont{and} \bibinfo{author}{\bibfnamefont{D.}~\bibnamefont{Kochan}}, \href{https://journals.aps.org/prb/abstract/10.1103/PhysRevB.108.054522}{\bibinfo{title}{Microscopic study of the josephson supercurrent diode effect in {Josephson} junctions based on two-dimensional electron gas}}, \bibinfo{journal}{Physical Review B} \textbf{\bibinfo{volume}{108}}, \bibinfo{pages}{054522} (\bibinfo{year}{2023}).

\bibitem[{\citenamefont{Maiani et~al.}(2023)\citenamefont{Maiani, Flensberg, Leijnse, Schrade, Vaitiek{\.e}nas, and Souto}}]{maiani2023nonsinusoidal}
\bibinfo{author}{\bibfnamefont{A.}~\bibnamefont{Maiani}}, \bibinfo{author}{\bibfnamefont{K.}~\bibnamefont{Flensberg}}, \bibinfo{author}{\bibfnamefont{M.}~\bibnamefont{Leijnse}}, \bibinfo{author}{\bibfnamefont{C.}~\bibnamefont{Schrade}}, \bibinfo{author}{\bibfnamefont{S.}~\bibnamefont{Vaitiek{\.e}nas}}, \bibnamefont{and} \bibinfo{author}{\bibfnamefont{R.~S.} \bibnamefont{Souto}}, \href{https://journals.aps.org/prb/abstract/10.1103/PhysRevB.107.245415}{\bibinfo{title}{Nonsinusoidal current-phase relations in semiconductor--superconductor--ferromagnetic insulator devices}}, \bibinfo{journal}{Physical Review B} \textbf{\bibinfo{volume}{107}}, \bibinfo{pages}{245415} (\bibinfo{year}{2023}).

\bibitem[{\citenamefont{Hess et~al.}(2023)\citenamefont{Hess, Legg, Loss, and Klinovaja}}]{hess2023josephson}
\bibinfo{author}{\bibfnamefont{R.}~\bibnamefont{Hess}}, \bibinfo{author}{\bibfnamefont{H.~F.} \bibnamefont{Legg}}, \bibinfo{author}{\bibfnamefont{D.}~\bibnamefont{Loss}}, \bibnamefont{and} \bibinfo{author}{\bibfnamefont{J.}~\bibnamefont{Klinovaja}}, \href{https://journals.aps.org/prb/abstract/10.1103/PhysRevB.108.174516}{\bibinfo{title}{Josephson transistor from the superconducting diode effect in domain wall and skyrmion magnetic racetracks}}, \bibinfo{journal}{Physical Review B} \textbf{\bibinfo{volume}{108}}, \bibinfo{pages}{174516} (\bibinfo{year}{2023}).

\bibitem[{\citenamefont{Hou et~al.}(2023)\citenamefont{Hou, Nichele, Chi, Lodesani, Wu, Ritter, Haxell, Davydova, Ili{\'c}, Glezakou-Elbert et~al.}}]{hou2023ubiquitous}
\bibinfo{author}{\bibfnamefont{Y.}~\bibnamefont{Hou}}, \bibinfo{author}{\bibfnamefont{F.}~\bibnamefont{Nichele}}, \bibinfo{author}{\bibfnamefont{H.}~\bibnamefont{Chi}}, \bibinfo{author}{\bibfnamefont{A.}~\bibnamefont{Lodesani}}, \bibinfo{author}{\bibfnamefont{Y.}~\bibnamefont{Wu}}, \bibinfo{author}{\bibfnamefont{M.~F.} \bibnamefont{Ritter}}, \bibinfo{author}{\bibfnamefont{D.~Z.} \bibnamefont{Haxell}}, \bibinfo{author}{\bibfnamefont{M.}~\bibnamefont{Davydova}}, \bibinfo{author}{\bibfnamefont{S.}~\bibnamefont{Ili{\'c}}}, \bibinfo{author}{\bibfnamefont{O.}~\bibnamefont{Glezakou-Elbert}}, \bibnamefont{et~al.}, \href{https://journals.aps.org/prl/abstract/10.1103/PhysRevLett.131.027001}{\bibinfo{title}{Ubiquitous superconducting diode effect in superconductor thin films}}, \bibinfo{journal}{Physical Review Letters} \textbf{\bibinfo{volume}{131}}, \bibinfo{pages}{027001} (\bibinfo{year}{2023}).

\bibitem[{\citenamefont{Kononov et~al.}(2020)\citenamefont{Kononov, Abulizi, Qu, Yan, Mandrus, Watanabe, Taniguchi, and Sch\"{o}nenberger}}]{kononov2020one}
\bibinfo{author}{\bibfnamefont{A.}~\bibnamefont{Kononov}}, \bibinfo{author}{\bibfnamefont{G.}~\bibnamefont{Abulizi}}, \bibinfo{author}{\bibfnamefont{K.}~\bibnamefont{Qu}}, \bibinfo{author}{\bibfnamefont{J.}~\bibnamefont{Yan}}, \bibinfo{author}{\bibfnamefont{D.}~\bibnamefont{Mandrus}}, \bibinfo{author}{\bibfnamefont{K.}~\bibnamefont{Watanabe}}, \bibinfo{author}{\bibfnamefont{T.}~\bibnamefont{Taniguchi}}, \bibnamefont{and} \bibinfo{author}{\bibfnamefont{C.}~\bibnamefont{Sch\"{o}nenberger}}, \href{https://pubs.acs.org/doi/10.1021/acs.nanolett.0c00658}{\bibinfo{title}{One-dimensional edge transport in few-layer $\text{WTe}_2$}}, \bibinfo{journal}{Nano letters} \textbf{\bibinfo{volume}{20}}, \bibinfo{pages}{4228} (\bibinfo{year}{2020}).

\bibitem[{\citenamefont{Souto et~al.}(2022)\citenamefont{Souto, Leijnse, and Schrade}}]{souto2022josephson}
\bibinfo{author}{\bibfnamefont{R.~S.} \bibnamefont{Souto}}, \bibinfo{author}{\bibfnamefont{M.}~\bibnamefont{Leijnse}}, \bibnamefont{and} \bibinfo{author}{\bibfnamefont{C.}~\bibnamefont{Schrade}}, \href{https://journals.aps.org/prl/abstract/10.1103/PhysRevLett.129.267702}{\bibinfo{title}{Josephson {Diode} {Effect} in {Supercurrent} {Interferometers}}}, \bibinfo{journal}{Physical Review Letters} \textbf{\bibinfo{volume}{129}}, \bibinfo{pages}{267702} (\bibinfo{year}{2022}).

\bibitem[{\citenamefont{Gupta et~al.}(2023)\citenamefont{Gupta, Graziano, Pendharkar, Dong, Dempsey, Palmstr{\o}m, and Pribiag}}]{gupta2023gate}
\bibinfo{author}{\bibfnamefont{M.}~\bibnamefont{Gupta}}, \bibinfo{author}{\bibfnamefont{G.~V.} \bibnamefont{Graziano}}, \bibinfo{author}{\bibfnamefont{M.}~\bibnamefont{Pendharkar}}, \bibinfo{author}{\bibfnamefont{J.~T.} \bibnamefont{Dong}}, \bibinfo{author}{\bibfnamefont{C.~P.} \bibnamefont{Dempsey}}, \bibinfo{author}{\bibfnamefont{C.}~\bibnamefont{Palmstr{\o}m}}, \bibnamefont{and} \bibinfo{author}{\bibfnamefont{V.~S.} \bibnamefont{Pribiag}}, \href{https://www.nature.com/articles/s41467-023-38856-0}{\bibinfo{title}{Gate-tunable superconducting diode effect in a three-terminal {Josephson} device}}, \bibinfo{journal}{Nature Communications} \textbf{\bibinfo{volume}{14}}, \bibinfo{pages}{3078} (\bibinfo{year}{2023}).

\bibitem[{\citenamefont{Ciaccia et~al.}(2023)\citenamefont{Ciaccia, Haller, Drachmann, Lindemann, Manfra, Schrade, and Sch\"{o}nenberger}}]{ciaccia2023gate}
\bibinfo{author}{\bibfnamefont{C.}~\bibnamefont{Ciaccia}}, \bibinfo{author}{\bibfnamefont{R.}~\bibnamefont{Haller}}, \bibinfo{author}{\bibfnamefont{A.~C.~C.} \bibnamefont{Drachmann}}, \bibinfo{author}{\bibfnamefont{T.}~\bibnamefont{Lindemann}}, \bibinfo{author}{\bibfnamefont{M.~J.} \bibnamefont{Manfra}}, \bibinfo{author}{\bibfnamefont{C.}~\bibnamefont{Schrade}}, \bibnamefont{and} \bibinfo{author}{\bibfnamefont{C.}~\bibnamefont{Sch\"onenberger}}, \href{https://link.aps.org/doi/10.1103/PhysRevResearch.5.033131}{\bibinfo{title}{Gate-tunable {Josephson} diode in proximitized {InAs} supercurrent interferometers}}, \bibinfo{journal}{Phys. Rev. Res.} \textbf{\bibinfo{volume}{5}}, \bibinfo{pages}{033131} (\bibinfo{year}{2023}).

\bibitem[{\citenamefont{Valentini et~al.}(2024)\citenamefont{Valentini, Sagi, Baghumyan, de~Gijsel, Jung, Calcaterra, Ballabio, Aguilera~Servin, Aggarwal, Janik et~al.}}]{valentini2023radio}
\bibinfo{author}{\bibfnamefont{M.}~\bibnamefont{Valentini}}, \bibinfo{author}{\bibfnamefont{O.}~\bibnamefont{Sagi}}, \bibinfo{author}{\bibfnamefont{L.}~\bibnamefont{Baghumyan}}, \bibinfo{author}{\bibfnamefont{T.}~\bibnamefont{de~Gijsel}}, \bibinfo{author}{\bibfnamefont{J.}~\bibnamefont{Jung}}, \bibinfo{author}{\bibfnamefont{S.}~\bibnamefont{Calcaterra}}, \bibinfo{author}{\bibfnamefont{A.}~\bibnamefont{Ballabio}}, \bibinfo{author}{\bibfnamefont{J.}~\bibnamefont{Aguilera~Servin}}, \bibinfo{author}{\bibfnamefont{K.}~\bibnamefont{Aggarwal}}, \bibinfo{author}{\bibfnamefont{M.}~\bibnamefont{Janik}}, \bibnamefont{et~al.}, \href{https://www.nature.com/articles/s41467-023-44114-0}{\bibinfo{title}{Parity-conserving {Cooper}-pair transport and ideal superconducting diode in planar germanium}}, \bibinfo{journal}{Nature Communications} \textbf{\bibinfo{volume}{15}}, \bibinfo{pages}{169} (\bibinfo{year}{2024}).

\bibitem[{\citenamefont{Greco et~al.}(2023)\citenamefont{Greco, Pichard, and Giazotto}}]{greco2023josephson}
\bibinfo{author}{\bibfnamefont{A.}~\bibnamefont{Greco}}, \bibinfo{author}{\bibfnamefont{Q.}~\bibnamefont{Pichard}}, \bibnamefont{and} \bibinfo{author}{\bibfnamefont{F.}~\bibnamefont{Giazotto}}, \href{https://pubs.aip.org/aip/apl/article/123/9/092601/2908308/Josephson-diode-effect-in-monolithic-dc-SQUIDs}{\bibinfo{title}{Josephson diode effect in monolithic dc-{SQUIDs} based on {3D} {Dayem} nanobridges}}, \bibinfo{journal}{Applied Physics Letters} \textbf{\bibinfo{volume}{123}} (\bibinfo{year}{2023}).

\bibitem[{\citenamefont{Legg et~al.}(2023)\citenamefont{Legg, Laubscher, Loss, and Klinovaja}}]{legg2023parity}
\bibinfo{author}{\bibfnamefont{H.~F.} \bibnamefont{Legg}}, \bibinfo{author}{\bibfnamefont{K.}~\bibnamefont{Laubscher}}, \bibinfo{author}{\bibfnamefont{D.}~\bibnamefont{Loss}}, \bibnamefont{and} \bibinfo{author}{\bibfnamefont{J.}~\bibnamefont{Klinovaja}}, \href{https://journals.aps.org/prb/abstract/10.1103/PhysRevB.108.214520}{\bibinfo{title}{Parity-protected superconducting diode effect in topological {Josephson} junctions}}, \bibinfo{journal}{Physical Review B} \textbf{\bibinfo{volume}{108}}, \bibinfo{pages}{214520} (\bibinfo{year}{2023}).

\bibitem[{\citenamefont{Cuozzo et~al.}(2024)\citenamefont{Cuozzo, Pan, Shabani, and Rossi}}]{cuozzo2023microwave}
\bibinfo{author}{\bibfnamefont{J.~J.} \bibnamefont{Cuozzo}}, \bibinfo{author}{\bibfnamefont{W.}~\bibnamefont{Pan}}, \bibinfo{author}{\bibfnamefont{J.}~\bibnamefont{Shabani}}, \bibnamefont{and} \bibinfo{author}{\bibfnamefont{E.}~\bibnamefont{Rossi}}, \href{https://journals.aps.org/prresearch/abstract/10.1103/PhysRevResearch.6.023011}{\bibinfo{title}{Microwave-tunable diode effect in asymmetric {SQUIDs} with topological {Josephson} junctions}}, \bibinfo{journal}{Physical Review Research} \textbf{\bibinfo{volume}{6}}, \bibinfo{pages}{023011} (\bibinfo{year}{2024}).

\bibitem[{\citenamefont{Lin et~al.}(2022)\citenamefont{Lin, Siriviboon, Scammell, Liu, Rhodes, Watanabe, Taniguchi, Hone, Scheurer, and Li}}]{ZeroFieldDiode}
\bibinfo{author}{\bibfnamefont{J.-X.} \bibnamefont{Lin}}, \bibinfo{author}{\bibfnamefont{P.}~\bibnamefont{Siriviboon}}, \bibinfo{author}{\bibfnamefont{H.~D.} \bibnamefont{Scammell}}, \bibinfo{author}{\bibfnamefont{S.}~\bibnamefont{Liu}}, \bibinfo{author}{\bibfnamefont{D.}~\bibnamefont{Rhodes}}, \bibinfo{author}{\bibfnamefont{K.}~\bibnamefont{Watanabe}}, \bibinfo{author}{\bibfnamefont{T.}~\bibnamefont{Taniguchi}}, \bibinfo{author}{\bibfnamefont{J.}~\bibnamefont{Hone}}, \bibinfo{author}{\bibfnamefont{M.~S.} \bibnamefont{Scheurer}}, \bibnamefont{and} \bibinfo{author}{\bibfnamefont{J.~I.~A.} \bibnamefont{Li}}, \href{https://doi.org/10.1038/s41567-022-01700-1}{\bibinfo{title}{Zero-field superconducting diode effect in small-twist-angle trilayer graphene}}, \bibinfo{journal}{Nature Physics} \textbf{\bibinfo{volume}{18}}, \bibinfo{pages}{1221} (\bibinfo{year}{2022}).

\bibitem[{\citenamefont{Scammell et~al.}(2022)\citenamefont{Scammell, Li, and Scheurer}}]{scammell_theory_2022}
\bibinfo{author}{\bibfnamefont{H.~D.} \bibnamefont{Scammell}}, \bibinfo{author}{\bibfnamefont{J.}~\bibnamefont{Li}}, \bibnamefont{and} \bibinfo{author}{\bibfnamefont{M.~S.} \bibnamefont{Scheurer}}, \href{https://iopscience.iop.org/article/10.1088/2053-1583/ac5b16}{\bibinfo{title}{Theory of zero-field superconducting diode effect in twisted trilayer graphene}}, \bibinfo{journal}{2D Materials} \textbf{\bibinfo{volume}{9}}, \bibinfo{pages}{025027} (\bibinfo{year}{2022}).

\bibitem[{\citenamefont{Wu et~al.}(2022)\citenamefont{Wu, Wang, Xu, Sivakumar, Pasco, Filippozzi, Parkin, Zeng, McQueen, and Ali}}]{wu_field-free_2022}
\bibinfo{author}{\bibfnamefont{H.}~\bibnamefont{Wu}}, \bibinfo{author}{\bibfnamefont{Y.}~\bibnamefont{Wang}}, \bibinfo{author}{\bibfnamefont{Y.}~\bibnamefont{Xu}}, \bibinfo{author}{\bibfnamefont{P.~K.} \bibnamefont{Sivakumar}}, \bibinfo{author}{\bibfnamefont{C.}~\bibnamefont{Pasco}}, \bibinfo{author}{\bibfnamefont{U.}~\bibnamefont{Filippozzi}}, \bibinfo{author}{\bibfnamefont{S.~S.~P.} \bibnamefont{Parkin}}, \bibinfo{author}{\bibfnamefont{Y.-J.} \bibnamefont{Zeng}}, \bibinfo{author}{\bibfnamefont{T.}~\bibnamefont{McQueen}}, \bibnamefont{and} \bibinfo{author}{\bibfnamefont{M.~N.} \bibnamefont{Ali}}, \href{https://www.nature.com/articles/s41586-022-04504-8}{\bibinfo{title}{The field-free {Josephson} diode in a van der {Waals} heterostructure}}, \bibinfo{journal}{Nature} \textbf{\bibinfo{volume}{604}}, \bibinfo{pages}{653} (\bibinfo{year}{2022}).

\bibitem[{\citenamefont{Zhang et~al.}(2024)\citenamefont{Zhang, Lin, Chichinadze, Wang, Watanabe, Taniguchi, Fu, and Li}}]{zhang2024angle}
\bibinfo{author}{\bibfnamefont{N.~J.} \bibnamefont{Zhang}}, \bibinfo{author}{\bibfnamefont{J.-X.} \bibnamefont{Lin}}, \bibinfo{author}{\bibfnamefont{D.~V.} \bibnamefont{Chichinadze}}, \bibinfo{author}{\bibfnamefont{Y.}~\bibnamefont{Wang}}, \bibinfo{author}{\bibfnamefont{K.}~\bibnamefont{Watanabe}}, \bibinfo{author}{\bibfnamefont{T.}~\bibnamefont{Taniguchi}}, \bibinfo{author}{\bibfnamefont{L.}~\bibnamefont{Fu}}, \bibnamefont{and} \bibinfo{author}{\bibfnamefont{J.}~\bibnamefont{Li}}, \href{https://www.nature.com/articles/s41563-024-01809-z}{\bibinfo{title}{Angle-resolved transport non-reciprocity and spontaneous symmetry breaking in twisted trilayer graphene}}, \bibinfo{journal}{Nature Materials} \textbf{\bibinfo{volume}{23}}, \bibinfo{pages}{356} (\bibinfo{year}{2024}).

\bibitem[{\citenamefont{Banerjee and Scheurer}(2024{\natexlab{a}})}]{PhysRevLett.132.046003}
\bibinfo{author}{\bibfnamefont{S.}~\bibnamefont{Banerjee}} \bibnamefont{and} \bibinfo{author}{\bibfnamefont{M.~S.} \bibnamefont{Scheurer}}, \href{https://link.aps.org/doi/10.1103/PhysRevLett.132.046003}{\bibinfo{title}{Enhanced superconducting diode effect due to coexisting phases}}, \bibinfo{journal}{Phys. Rev. Lett.} \textbf{\bibinfo{volume}{132}}, \bibinfo{pages}{046003} (\bibinfo{year}{2024}{\natexlab{a}}).

\bibitem[{\citenamefont{D{\'\i}ez-M{\'e}rida et~al.}(2023)\citenamefont{D{\'\i}ez-M{\'e}rida, D{\'\i}ez-Carl{\'o}n, Yang, Xie, Gao, Senior, Watanabe, Taniguchi, Lu, Higginbotham et~al.}}]{diez2023symmetry}
\bibinfo{author}{\bibfnamefont{J.}~\bibnamefont{D{\'\i}ez-M{\'e}rida}}, \bibinfo{author}{\bibfnamefont{A.}~\bibnamefont{D{\'\i}ez-Carl{\'o}n}}, \bibinfo{author}{\bibfnamefont{S.}~\bibnamefont{Yang}}, \bibinfo{author}{\bibfnamefont{Y.-M.} \bibnamefont{Xie}}, \bibinfo{author}{\bibfnamefont{X.-J.} \bibnamefont{Gao}}, \bibinfo{author}{\bibfnamefont{J.}~\bibnamefont{Senior}}, \bibinfo{author}{\bibfnamefont{K.}~\bibnamefont{Watanabe}}, \bibinfo{author}{\bibfnamefont{T.}~\bibnamefont{Taniguchi}}, \bibinfo{author}{\bibfnamefont{X.}~\bibnamefont{Lu}}, \bibinfo{author}{\bibfnamefont{A.~P.} \bibnamefont{Higginbotham}}, \bibnamefont{et~al.}, \href{https://www.nature.com/articles/s41467-023-38005-7}{\bibinfo{title}{Symmetry-broken {Josephson} junctions and superconducting diodes in magic-angle twisted bilayer graphene}}, \bibinfo{journal}{Nature Communications} \textbf{\bibinfo{volume}{14}}, \bibinfo{pages}{2396} (\bibinfo{year}{2023}).

\bibitem[{\citenamefont{Hu et~al.}(2023)\citenamefont{Hu, Sun, Xie, and Law}}]{hu2023josephson}
\bibinfo{author}{\bibfnamefont{J.-X.} \bibnamefont{Hu}}, \bibinfo{author}{\bibfnamefont{Z.-T.} \bibnamefont{Sun}}, \bibinfo{author}{\bibfnamefont{Y.-M.} \bibnamefont{Xie}}, \bibnamefont{and} \bibinfo{author}{\bibfnamefont{K.~T.} \bibnamefont{Law}}, \href{https://journals.aps.org/prl/abstract/10.1103/PhysRevLett.130.266003}{\bibinfo{title}{Josephson {Diode} {Effect} {Induced} by {Valley} {Polarization} in {Twisted} {Bilayer} {Graphene}}}, \bibinfo{journal}{Physical Review Letters} \textbf{\bibinfo{volume}{130}}, \bibinfo{pages}{266003} (\bibinfo{year}{2023}).

\bibitem[{\citenamefont{Khabipov et~al.}(2022)\citenamefont{Khabipov, Gaydamachenko, Kissling, Dolata, and Zorin}}]{khabipov2022superconducting}
\bibinfo{author}{\bibfnamefont{M.}~\bibnamefont{Khabipov}}, \bibinfo{author}{\bibfnamefont{V.}~\bibnamefont{Gaydamachenko}}, \bibinfo{author}{\bibfnamefont{C.}~\bibnamefont{Kissling}}, \bibinfo{author}{\bibfnamefont{R.}~\bibnamefont{Dolata}}, \bibnamefont{and} \bibinfo{author}{\bibfnamefont{A.}~\bibnamefont{Zorin}}, \href{https://iopscience.iop.org/article/10.1088/1361-6668/ac6989}{\bibinfo{title}{Superconducting microwave resonators with non-centrosymmetric nonlinearity}}, \bibinfo{journal}{Superconductor Science and Technology} \textbf{\bibinfo{volume}{35}}, \bibinfo{pages}{065020} (\bibinfo{year}{2022}).

\bibitem[{\citenamefont{Frattini et~al.}(2017)\citenamefont{Frattini, Vool, Shankar, Narla, Sliwa, and Devoret}}]{frattini_3-wave_2017}
\bibinfo{author}{\bibfnamefont{N.~E.} \bibnamefont{Frattini}}, \bibinfo{author}{\bibfnamefont{U.}~\bibnamefont{Vool}}, \bibinfo{author}{\bibfnamefont{S.}~\bibnamefont{Shankar}}, \bibinfo{author}{\bibfnamefont{A.}~\bibnamefont{Narla}}, \bibinfo{author}{\bibfnamefont{K.~M.} \bibnamefont{Sliwa}}, \bibnamefont{and} \bibinfo{author}{\bibfnamefont{M.~H.} \bibnamefont{Devoret}}, \href{https://aip.scitation.org/doi/10.1063/1.4984142}{\bibinfo{title}{3-wave mixing {Josephson} dipole element}}, \bibinfo{journal}{Applied Physics Letters} \textbf{\bibinfo{volume}{110}}, \bibinfo{pages}{222603} (\bibinfo{year}{2017}).

\bibitem[{\citenamefont{Frattini et~al.}(2018)\citenamefont{Frattini, Sivak, Lingenfelter, Shankar, and Devoret}}]{frattini_optimizing_2018}
\bibinfo{author}{\bibfnamefont{N.}~\bibnamefont{Frattini}}, \bibinfo{author}{\bibfnamefont{V.}~\bibnamefont{Sivak}}, \bibinfo{author}{\bibfnamefont{A.}~\bibnamefont{Lingenfelter}}, \bibinfo{author}{\bibfnamefont{S.}~\bibnamefont{Shankar}}, \bibnamefont{and} \bibinfo{author}{\bibfnamefont{M.}~\bibnamefont{Devoret}}, \href{https://journals.aps.org/prapplied/abstract/10.1103/PhysRevApplied.10.054020}{\bibinfo{title}{Optimizing the nonlinearity and dissipation of a snail parametric amplifier for dynamic range}}, \bibinfo{journal}{Physical Review Applied} \textbf{\bibinfo{volume}{10}}, \bibinfo{pages}{054020} (\bibinfo{year}{2018}).

\bibitem[{\citenamefont{Sivak et~al.}(2019)\citenamefont{Sivak, Frattini, Joshi, Lingenfelter, Shankar, and Devoret}}]{sivak_kerr-free_2019}
\bibinfo{author}{\bibfnamefont{V.}~\bibnamefont{Sivak}}, \bibinfo{author}{\bibfnamefont{N.}~\bibnamefont{Frattini}}, \bibinfo{author}{\bibfnamefont{V.}~\bibnamefont{Joshi}}, \bibinfo{author}{\bibfnamefont{A.}~\bibnamefont{Lingenfelter}}, \bibinfo{author}{\bibfnamefont{S.}~\bibnamefont{Shankar}}, \bibnamefont{and} \bibinfo{author}{\bibfnamefont{M.}~\bibnamefont{Devoret}}, \href{https://link.aps.org/doi/10.1103/PhysRevApplied.11.054060}{\bibinfo{title}{Kerr-{Free} {Three}-{Wave} {Mixing} in {Superconducting} {Quantum} {Circuits}}}, \bibinfo{journal}{Physical Review Applied} \textbf{\bibinfo{volume}{11}}, \bibinfo{pages}{054060} (\bibinfo{year}{2019}).

\bibitem[{\citenamefont{Miano et~al.}(2022)\citenamefont{Miano, Liu, Sivak, Frattini, Joshi, Dai, Frunzio, and Devoret}}]{miano_frequency-tunable_2022}
\bibinfo{author}{\bibfnamefont{A.}~\bibnamefont{Miano}}, \bibinfo{author}{\bibfnamefont{G.}~\bibnamefont{Liu}}, \bibinfo{author}{\bibfnamefont{V.~V.} \bibnamefont{Sivak}}, \bibinfo{author}{\bibfnamefont{N.~E.} \bibnamefont{Frattini}}, \bibinfo{author}{\bibfnamefont{V.~R.} \bibnamefont{Joshi}}, \bibinfo{author}{\bibfnamefont{W.}~\bibnamefont{Dai}}, \bibinfo{author}{\bibfnamefont{L.}~\bibnamefont{Frunzio}}, \bibnamefont{and} \bibinfo{author}{\bibfnamefont{M.~H.} \bibnamefont{Devoret}}, \href{https://doi.org/10.1063/5.0083350}{\bibinfo{title}{Frequency-tunable {Kerr}-free three-wave mixing with a gradiometric {SNAIL}}}, \bibinfo{journal}{Applied Physics Letters} \textbf{\bibinfo{volume}{120}}, \bibinfo{pages}{184002} (\bibinfo{year}{2022}).

\bibitem[{\citenamefont{Schrade and Fatemi}(2024)}]{PhysRevApplied.21.064029}
\bibinfo{author}{\bibfnamefont{C.}~\bibnamefont{Schrade}} \bibnamefont{and} \bibinfo{author}{\bibfnamefont{V.}~\bibnamefont{Fatemi}}, \href{https://link.aps.org/doi/10.1103/PhysRevApplied.21.064029}{\bibinfo{title}{Dissipationless nonlinearity in quantum material josephson diodes}}, \bibinfo{journal}{Phys. Rev. Appl.} \textbf{\bibinfo{volume}{21}}, \bibinfo{pages}{064029} (\bibinfo{year}{2024}).

\bibitem[{\citenamefont{Han et~al.}(2024)\citenamefont{Han, Lu, Hadjri, Shi, Wu, Xu, Yao, Cotten, Sedeh, Weldeyesus et~al.}}]{han2024}
\bibinfo{author}{\bibfnamefont{T.}~\bibnamefont{Han}}, \bibinfo{author}{\bibfnamefont{Z.}~\bibnamefont{Lu}}, \bibinfo{author}{\bibfnamefont{Z.}~\bibnamefont{Hadjri}}, \bibinfo{author}{\bibfnamefont{L.}~\bibnamefont{Shi}}, \bibinfo{author}{\bibfnamefont{Z.}~\bibnamefont{Wu}}, \bibinfo{author}{\bibfnamefont{W.}~\bibnamefont{Xu}}, \bibinfo{author}{\bibfnamefont{Y.}~\bibnamefont{Yao}}, \bibinfo{author}{\bibfnamefont{A.~A.} \bibnamefont{Cotten}}, \bibinfo{author}{\bibfnamefont{O.~S.} \bibnamefont{Sedeh}}, \bibinfo{author}{\bibfnamefont{H.}~\bibnamefont{Weldeyesus}}, \bibnamefont{et~al.}, \href{https://arxiv.org/abs/2408.15233}{\bibinfo{title}{Signatures of chiral superconductivity in rhombohedral graphene}}, \bibinfo{journal}{arXiv preprint arXiv:2408.15233}  (\bibinfo{year}{2024}).

\bibitem[{\citenamefont{Choi et~al.}(2024)\citenamefont{Choi, Choi, Valentini, Patterson, Holleis, Sheekey, Stoyanov, Cheng, Taniguchi, Watanabe et~al.}}]{choi2024}
\bibinfo{author}{\bibfnamefont{Y.}~\bibnamefont{Choi}}, \bibinfo{author}{\bibfnamefont{Y.}~\bibnamefont{Choi}}, \bibinfo{author}{\bibfnamefont{M.}~\bibnamefont{Valentini}}, \bibinfo{author}{\bibfnamefont{C.~L.} \bibnamefont{Patterson}}, \bibinfo{author}{\bibfnamefont{L.~F.~W.} \bibnamefont{Holleis}}, \bibinfo{author}{\bibfnamefont{O.~I.} \bibnamefont{Sheekey}}, \bibinfo{author}{\bibfnamefont{H.}~\bibnamefont{Stoyanov}}, \bibinfo{author}{\bibfnamefont{X.}~\bibnamefont{Cheng}}, \bibinfo{author}{\bibfnamefont{T.}~\bibnamefont{Taniguchi}}, \bibinfo{author}{\bibfnamefont{K.}~\bibnamefont{Watanabe}}, \bibnamefont{et~al.}, \href{https://arxiv.org/abs/2408.12584}{\bibinfo{title}{Electric field control of superconductivity and quantized anomalous hall effects in rhombohedral tetralayer graphene}}, \bibinfo{journal}{arXiv preprint arXiv:2408.12584}  (\bibinfo{year}{2024}).

\bibitem[{\citenamefont{Geier et~al.}(2024)\citenamefont{Geier, Davydova, and Fu}}]{geier2024chiral}
\bibinfo{author}{\bibfnamefont{M.}~\bibnamefont{Geier}}, \bibinfo{author}{\bibfnamefont{M.}~\bibnamefont{Davydova}}, \bibnamefont{and} \bibinfo{author}{\bibfnamefont{L.}~\bibnamefont{Fu}}, \href{https://arxiv.org/abs/2409.13829}{\bibinfo{title}{Chiral and topological superconductivity in isospin polarized multilayer graphene}}, \bibinfo{journal}{arXiv preprint arXiv:2409.13829}  (\bibinfo{year}{2024}).

\bibitem[{\citenamefont{Chou et~al.}(2024)\citenamefont{Chou, Zhu, and Sarma}}]{chou2024intravalley}
\bibinfo{author}{\bibfnamefont{Y.-Z.} \bibnamefont{Chou}}, \bibinfo{author}{\bibfnamefont{J.}~\bibnamefont{Zhu}}, \bibnamefont{and} \bibinfo{author}{\bibfnamefont{S.~D.} \bibnamefont{Sarma}}, \href{https://arxiv.org/abs/2409.06701}{\bibinfo{title}{Intravalley spin-polarized superconductivity in rhombohedral tetralayer graphene}}, \bibinfo{journal}{arXiv preprint arXiv:2409.06701}  (\bibinfo{year}{2024}).

\bibitem[{\citenamefont{Yang and Zhang}(2024)}]{yang2024topological}
\bibinfo{author}{\bibfnamefont{H.}~\bibnamefont{Yang}} \bibnamefont{and} \bibinfo{author}{\bibfnamefont{Y.-H.} \bibnamefont{Zhang}}, \href{https://arxiv.org/abs/2411.02503}{\bibinfo{title}{Topological incommensurate fulde-ferrell-larkin-ovchinnikov superconductor and bogoliubov fermi surface in rhombohedral tetra-layer graphene}}, \bibinfo{journal}{arXiv preprint arXiv:2411.02503}  (\bibinfo{year}{2024}).

\bibitem[{\citenamefont{Qin and Wu}(2024)}]{qin2024chiral}
\bibinfo{author}{\bibfnamefont{Q.}~\bibnamefont{Qin}} \bibnamefont{and} \bibinfo{author}{\bibfnamefont{C.}~\bibnamefont{Wu}}, \href{https://arxiv.org/abs/2412.07145}{\bibinfo{title}{Chiral finite-momentum superconductivity in the tetralayer graphene}}, \bibinfo{journal}{arXiv preprint arXiv:2412.07145}  (\bibinfo{year}{2024}).

\bibitem[{\citenamefont{Parra-Martinez et~al.}(2025)\citenamefont{Parra-Martinez, Jimeno-Pozo, Phong, Sainz-Cruz, Kaplan, Emanuel, Oreg, Pantale{\'o}n, Silva-Guill{\'e}n, and Guinea}}]{parra2025band}
\bibinfo{author}{\bibfnamefont{G.}~\bibnamefont{Parra-Martinez}}, \bibinfo{author}{\bibfnamefont{A.}~\bibnamefont{Jimeno-Pozo}}, \bibinfo{author}{\bibfnamefont{V.~T.} \bibnamefont{Phong}}, \bibinfo{author}{\bibfnamefont{H.}~\bibnamefont{Sainz-Cruz}}, \bibinfo{author}{\bibfnamefont{D.}~\bibnamefont{Kaplan}}, \bibinfo{author}{\bibfnamefont{P.}~\bibnamefont{Emanuel}}, \bibinfo{author}{\bibfnamefont{Y.}~\bibnamefont{Oreg}}, \bibinfo{author}{\bibfnamefont{P.~A.} \bibnamefont{Pantale{\'o}n}}, \bibinfo{author}{\bibfnamefont{J.~A.} \bibnamefont{Silva-Guill{\'e}n}}, \bibnamefont{and} \bibinfo{author}{\bibfnamefont{F.}~\bibnamefont{Guinea}}, \href{https://arxiv.org/abs/2502.19474}{\bibinfo{title}{Band renormalization, quarter metals, and chiral superconductivity in rhombohedral tetralayer graphene}}, \bibinfo{journal}{arXiv preprint arXiv:2502.19474}  (\bibinfo{year}{2025}).

\bibitem[{\citenamefont{Dong and Lee}(2025)}]{dong2025}
\bibinfo{author}{\bibfnamefont{Z.}~\bibnamefont{Dong}} \bibnamefont{and} \bibinfo{author}{\bibfnamefont{P.~A.} \bibnamefont{Lee}}, \href{https://arxiv.org/abs/2503.11079}{\bibinfo{title}{A controllable theory of superconductivity due to strong repulsion in a polarized band}} (\bibinfo{year}{2025}), \eprint{2503.11079}.

\bibitem[{\citenamefont{Jahin and Lin}(2024)}]{jahin2024enhanced}
\bibinfo{author}{\bibfnamefont{A.}~\bibnamefont{Jahin}} \bibnamefont{and} \bibinfo{author}{\bibfnamefont{S.-Z.} \bibnamefont{Lin}}, \href{https://arxiv.org/abs/2411.09664}{\bibinfo{title}{Enhanced kohn-luttinger topological superconductivity in bands with nontrivial geometry}}, \bibinfo{journal}{arXiv preprint arXiv:2411.09664}  (\bibinfo{year}{2024}).

\bibitem[{\citenamefont{Wang et~al.}(2024)\citenamefont{Wang, Gao, and Yang}}]{wang2024chiral}
\bibinfo{author}{\bibfnamefont{Y.-Q.} \bibnamefont{Wang}}, \bibinfo{author}{\bibfnamefont{Z.-Q.} \bibnamefont{Gao}}, \bibnamefont{and} \bibinfo{author}{\bibfnamefont{H.}~\bibnamefont{Yang}}, \href{https://arxiv.org/abs/2410.05384}{\bibinfo{title}{Chiral superconductivity from parent chern band and its non-abelian generalization}}, \bibinfo{journal}{arXiv preprint arXiv:2410.05384}  (\bibinfo{year}{2024}).

\bibitem[{\citenamefont{Yoon et~al.}(2025)\citenamefont{Yoon, Xu, Barlas, and Zhang}}]{yoon2025quarter}
\bibinfo{author}{\bibfnamefont{C.}~\bibnamefont{Yoon}}, \bibinfo{author}{\bibfnamefont{T.}~\bibnamefont{Xu}}, \bibinfo{author}{\bibfnamefont{Y.}~\bibnamefont{Barlas}}, \bibnamefont{and} \bibinfo{author}{\bibfnamefont{F.}~\bibnamefont{Zhang}}, \href{https://arxiv.org/abs/2502.17555}{\bibinfo{title}{Quarter metal superconductivity}}, \bibinfo{journal}{arXiv preprint arXiv:2502.17555}  (\bibinfo{year}{2025}).

\bibitem[{\citenamefont{May-Mann et~al.}(2025)\citenamefont{May-Mann, Helbig, and Devakul}}]{maymann2025}
\bibinfo{author}{\bibfnamefont{J.}~\bibnamefont{May-Mann}}, \bibinfo{author}{\bibfnamefont{T.}~\bibnamefont{Helbig}}, \bibnamefont{and} \bibinfo{author}{\bibfnamefont{T.}~\bibnamefont{Devakul}}, \href{https://arxiv.org/abs/2503.05697}{\bibinfo{title}{How pairing mechanism dictates topology in valley-polarized superconductors with berry curvature}} (\bibinfo{year}{2025}), \eprint{2503.05697}.

\bibitem[{\citenamefont{Christos et~al.}(2025)\citenamefont{Christos, Bonetti, and Scheurer}}]{christos2025}
\bibinfo{author}{\bibfnamefont{M.}~\bibnamefont{Christos}}, \bibinfo{author}{\bibfnamefont{P.~M.} \bibnamefont{Bonetti}}, \bibnamefont{and} \bibinfo{author}{\bibfnamefont{M.~S.} \bibnamefont{Scheurer}}, \href{https://arxiv.org/abs/2503.15471}{\bibinfo{title}{Finite-momentum pairing and superlattice superconductivity in valley-imbalanced rhombohedral graphene}} (\bibinfo{year}{2025}), \eprint{2503.15471}.

\bibitem[{\citenamefont{Koshino and McCann}(2009)}]{Model_Koshino}
\bibinfo{author}{\bibfnamefont{M.}~\bibnamefont{Koshino}} \bibnamefont{and} \bibinfo{author}{\bibfnamefont{E.}~\bibnamefont{McCann}}, \href{https://link.aps.org/doi/10.1103/PhysRevB.80.165409}{\bibinfo{title}{Trigonal warping and berry's phase $n\ensuremath{\pi}$ in abc-stacked multilayer graphene}}, \bibinfo{journal}{Phys. Rev. B} \textbf{\bibinfo{volume}{80}}, \bibinfo{pages}{165409} (\bibinfo{year}{2009}).

\bibitem[{\citenamefont{Zhang et~al.}(2010)\citenamefont{Zhang, Sahu, Min, and MacDonald}}]{PhysRevB.82.035409}
\bibinfo{author}{\bibfnamefont{F.}~\bibnamefont{Zhang}}, \bibinfo{author}{\bibfnamefont{B.}~\bibnamefont{Sahu}}, \bibinfo{author}{\bibfnamefont{H.}~\bibnamefont{Min}}, \bibnamefont{and} \bibinfo{author}{\bibfnamefont{A.~H.} \bibnamefont{MacDonald}}, \href{https://link.aps.org/doi/10.1103/PhysRevB.82.035409}{\bibinfo{title}{Band structure of $abc$-stacked graphene trilayers}}, \bibinfo{journal}{Phys. Rev. B} \textbf{\bibinfo{volume}{82}}, \bibinfo{pages}{035409} (\bibinfo{year}{2010}).

\bibitem[{\citenamefont{Ghazaryan et~al.}(2023)\citenamefont{Ghazaryan, Holder, Berg, and Serbyn}}]{PhysRevB.107.104502}
\bibinfo{author}{\bibfnamefont{A.}~\bibnamefont{Ghazaryan}}, \bibinfo{author}{\bibfnamefont{T.}~\bibnamefont{Holder}}, \bibinfo{author}{\bibfnamefont{E.}~\bibnamefont{Berg}}, \bibnamefont{and} \bibinfo{author}{\bibfnamefont{M.}~\bibnamefont{Serbyn}}, \href{https://link.aps.org/doi/10.1103/PhysRevB.107.104502}{\bibinfo{title}{Multilayer graphenes as a platform for interaction-driven physics and topological superconductivity}}, \bibinfo{journal}{Phys. Rev. B} \textbf{\bibinfo{volume}{107}}, \bibinfo{pages}{104502} (\bibinfo{year}{2023}).

\bibitem[{\citenamefont{Zhou et~al.}(2021)\citenamefont{Zhou, Xie, Ghazaryan, Holder, Ehrets, Spanton, Taniguchi, Watanabe, Berg, Serbyn et~al.}}]{zhou2021half}
\bibinfo{author}{\bibfnamefont{H.}~\bibnamefont{Zhou}}, \bibinfo{author}{\bibfnamefont{T.}~\bibnamefont{Xie}}, \bibinfo{author}{\bibfnamefont{A.}~\bibnamefont{Ghazaryan}}, \bibinfo{author}{\bibfnamefont{T.}~\bibnamefont{Holder}}, \bibinfo{author}{\bibfnamefont{J.~R.} \bibnamefont{Ehrets}}, \bibinfo{author}{\bibfnamefont{E.~M.} \bibnamefont{Spanton}}, \bibinfo{author}{\bibfnamefont{T.}~\bibnamefont{Taniguchi}}, \bibinfo{author}{\bibfnamefont{K.}~\bibnamefont{Watanabe}}, \bibinfo{author}{\bibfnamefont{E.}~\bibnamefont{Berg}}, \bibinfo{author}{\bibfnamefont{M.}~\bibnamefont{Serbyn}}, \bibnamefont{et~al.}, \href{https://www.nature.com/articles/s41586-021-03938-w}{\bibinfo{title}{Half-and quarter-metals in rhombohedral trilayer graphene}}, \bibinfo{journal}{Nature} \textbf{\bibinfo{volume}{598}}, \bibinfo{pages}{429} (\bibinfo{year}{2021}).

\bibitem[{\citenamefont{Davydova et~al.}(2024)\citenamefont{Davydova, Geier, and Fu}}]{davydovageier2024}
\bibinfo{author}{\bibfnamefont{M.}~\bibnamefont{Davydova}}, \bibinfo{author}{\bibfnamefont{M.}~\bibnamefont{Geier}}, \bibnamefont{and} \bibinfo{author}{\bibfnamefont{L.}~\bibnamefont{Fu}}, \href{https://www.science.org/doi/abs/10.1126/sciadv.adr4817}{\bibinfo{title}{Nonreciprocal superconductivity}}, \bibinfo{journal}{Science Advances} \textbf{\bibinfo{volume}{10}}, \bibinfo{pages}{eadr4817} (\bibinfo{year}{2024}), \eprint{https://www.science.org/doi/pdf/10.1126/sciadv.adr4817}.

\bibitem[{\citenamefont{Kohn and Luttinger}(1965)}]{kohn1965new}
\bibinfo{author}{\bibfnamefont{W.}~\bibnamefont{Kohn}} \bibnamefont{and} \bibinfo{author}{\bibfnamefont{J.}~\bibnamefont{Luttinger}}, \href{https://journals.aps.org/prl/abstract/10.1103/PhysRevLett.15.524}{\bibinfo{title}{New mechanism for superconductivity}}, \bibinfo{journal}{Physical Review Letters} \textbf{\bibinfo{volume}{15}}, \bibinfo{pages}{524} (\bibinfo{year}{1965}).

\bibitem[{\citenamefont{Chubukov}(1993)}]{chubukov1993kohn}
\bibinfo{author}{\bibfnamefont{A.~V.} \bibnamefont{Chubukov}}, \href{https://journals.aps.org/prb/abstract/10.1103/PhysRevB.48.1097}{\bibinfo{title}{Kohn-luttinger effect and the instability of a two-dimensional repulsive fermi liquid at t= 0}}, \bibinfo{journal}{Physical Review B} \textbf{\bibinfo{volume}{48}}, \bibinfo{pages}{1097} (\bibinfo{year}{1993}).

\bibitem[{\citenamefont{Ghazaryan et~al.}(2021)\citenamefont{Ghazaryan, Holder, Serbyn, and Berg}}]{PhysRevLett.127.247001}
\bibinfo{author}{\bibfnamefont{A.}~\bibnamefont{Ghazaryan}}, \bibinfo{author}{\bibfnamefont{T.}~\bibnamefont{Holder}}, \bibinfo{author}{\bibfnamefont{M.}~\bibnamefont{Serbyn}}, \bibnamefont{and} \bibinfo{author}{\bibfnamefont{E.}~\bibnamefont{Berg}}, \href{https://link.aps.org/doi/10.1103/PhysRevLett.127.247001}{\bibinfo{title}{Unconventional superconductivity in systems with annular fermi surfaces: Application to rhombohedral trilayer graphene}}, \bibinfo{journal}{Phys. Rev. Lett.} \textbf{\bibinfo{volume}{127}}, \bibinfo{pages}{247001} (\bibinfo{year}{2021}).

\bibitem[{\citenamefont{Schrade and Fu}(2024)}]{PhysRevB.110.035143}
\bibinfo{author}{\bibfnamefont{C.}~\bibnamefont{Schrade}} \bibnamefont{and} \bibinfo{author}{\bibfnamefont{L.}~\bibnamefont{Fu}}, \href{https://link.aps.org/doi/10.1103/PhysRevB.110.035143}{\bibinfo{title}{Nematic, chiral, and topological superconductivity in twisted transition metal dichalcogenides}}, \bibinfo{journal}{Phys. Rev. B} \textbf{\bibinfo{volume}{110}}, \bibinfo{pages}{035143} (\bibinfo{year}{2024}).

\bibitem[{\citenamefont{Daido et~al.}(2022)\citenamefont{Daido, Ikeda, and Yanase}}]{daido2022intrinsic}
\bibinfo{author}{\bibfnamefont{A.}~\bibnamefont{Daido}}, \bibinfo{author}{\bibfnamefont{Y.}~\bibnamefont{Ikeda}}, \bibnamefont{and} \bibinfo{author}{\bibfnamefont{Y.}~\bibnamefont{Yanase}}, \href{https://journals.aps.org/prl/abstract/10.1103/PhysRevLett.128.037001}{\bibinfo{title}{Intrinsic superconducting diode effect}}, \bibinfo{journal}{Physical Review Letters} \textbf{\bibinfo{volume}{128}}, \bibinfo{pages}{037001} (\bibinfo{year}{2022}).

\bibitem[{Sup()}]{Supplemental}
\bibinfo{note}{In the Supplemental Material, we provide details on the angular dependence of the diode efficiency and the the mean-field approach for describing the superconducting RTLG system.}

\bibitem[{\citenamefont{Banerjee and Scheurer}(2024{\natexlab{b}})}]{PhysRevB.110.024503}
\bibinfo{author}{\bibfnamefont{S.}~\bibnamefont{Banerjee}} \bibnamefont{and} \bibinfo{author}{\bibfnamefont{M.~S.} \bibnamefont{Scheurer}}, \href{https://link.aps.org/doi/10.1103/PhysRevB.110.024503}{\bibinfo{title}{Altermagnetic superconducting diode effect}}, \bibinfo{journal}{Phys. Rev. B} \textbf{\bibinfo{volume}{110}}, \bibinfo{pages}{024503} (\bibinfo{year}{2024}{\natexlab{b}}).

\bibitem[{\citenamefont{{Morissette} et~al.}(2025)\citenamefont{{Morissette}, {Qin}, {Wu}, {Zhang}, {Watanabe}, {Taniguchi}, and {Li}}}]{JiaNew}
\bibinfo{author}{\bibfnamefont{E.}~\bibnamefont{{Morissette}}}, \bibinfo{author}{\bibfnamefont{P.}~\bibnamefont{{Qin}}}, \bibinfo{author}{\bibfnamefont{H.-T.} \bibnamefont{{Wu}}}, \bibinfo{author}{\bibfnamefont{N.~J.} \bibnamefont{{Zhang}}}, \bibinfo{author}{\bibfnamefont{K.}~\bibnamefont{{Watanabe}}}, \bibinfo{author}{\bibfnamefont{T.}~\bibnamefont{{Taniguchi}}}, \bibnamefont{and} \bibinfo{author}{\bibfnamefont{J.~I.~A.} \bibnamefont{{Li}}}, \href{https://arxiv.org/abs/2504.05129}{\bibinfo{title}{{Superconductivity, Anomalous Hall Effect, and Stripe Order in Rhombohedral Hexalayer Graphene}}}, \bibinfo{journal}{arXiv e-prints}  (\bibinfo{year}{2025}), \eprint{2504.05129}.

\bibitem[{\citenamefont{Sedov and Scheurer}(2025)}]{sedov2025}
\bibinfo{author}{\bibfnamefont{D.}~\bibnamefont{Sedov}} \bibnamefont{and} \bibinfo{author}{\bibfnamefont{M.~S.} \bibnamefont{Scheurer}}, \href{https://arxiv.org/abs/2503.12650}{\bibinfo{title}{Probing superconductivity with tunneling spectroscopy in rhombohedral graphene}} (\bibinfo{year}{2025}), \eprint{2503.12650}.

\end{thebibliography}
\end{document}